%% file: paper.tex
\documentclass[sigconf]{acmart}

\usepackage{tabto}
\usepackage{hyperref}
\usepackage{array}
\usepackage{xcolor}
\usepackage{graphicx}
\usepackage{listings}
\usepackage{paralist}
\usepackage{url}
\usepackage{framed}
\definecolor{pdfbgcolor}{RGB}{180,180,180}

\newcommand{\fakeparagraph}[1]{\vspace{1mm}\noindent\textbf{#1.}}

\newcommand{\wcnarrow}[2]{\parbox{\namewidth}{#1} \com \mbox{#2}}

\newcommand{\comnospace}{\mbox{$\triangleright$}}
\newcommand{\com}{\mbox{\comnospace\ }}

\newcommand{\func}[1]{\mbox{\sc #1}}

\newcommand{\cas}{\mbox{CAS}}

\newcommand{\true}{\textsc{True}}
\newcommand{\false}{\textsc{False}}

\newcommand{\op}{\mbox{\sct -record}}
\newcommand{\rec}{\mbox{Data-record}}
\newcommand{\llresults}{\info Fields}
\newcommand{\info}{\textit{info}}

\newcommand{\help}{\func{Help}}
\newcommand{\llt}{\func{LLX}}
\newcommand{\sct}{\func{SCX}}

\newcommand{\freezing}{\mbox{InProgress}}
\newcommand{\retry}{\mbox{Aborted}}
\newcommand{\done}{\mbox{Committed}}

\newcommand{\freezingdone}{allFrozen}

\newcommand{\fail}{\textsc{Fail}}
\newcommand{\finalized}{\textsc{Finalized}}

\newcommand{\fcas}{{freezing~\cas}}
\newcommand{\astep}{{abort step}}
\newcommand{\fstep}{{frozen step}}
\newcommand{\fcstep}{{frozen check step}}
\newcommand{\cstep}{{commit step}}
\newcommand{\upcas}{{update~\cas}}
\newcommand{\markstep}{{mark step}}

\newcommand{\nil}{\textsc{Nil}}

\def\pwidth{2.5cm}
\def\indentamt{0mm}

\newcommand{\sidecom}[1]{\tabto{12.3cm} \com \mbox{#1}}

\newcommand{\dline}[2]{#1\\ \mbox{\hspace{\indentamt}#2}}
\newcommand{\tline}[3]{\dline{#1}{#2}\\ \mbox{\hspace{\indentamt}#3}}

\newcommand{\presctlinked}{1}
\newcommand{\presctabainit}{2}
\newcommand{\presctaba}{3}

\lstdefinestyle{nonumbers}{numbers=none}
\newcommand{\preplisting}{\lstset{gobble=1, numbers=left, numberstyle=\tiny, numberblanklines=false, firstnumber=last, numbersep=4pt, escapeinside={//}{\^^M}, breaklines=true, keywordstyle=\bfseries, morekeywords={type,subtype,break,continue,if,else,end,loop,while,do,done,exit, when,then,return,read,and,or,not,for,each,boolean,procedure,invoke,next,iteration,until,goto}}}
\newcommand{\prepnewlisting}{\lstset{gobble=1, numbers=left, numberstyle=\tiny, numberblanklines=false, firstnumber=1, numbersep=3pt, escapeinside={//}{\^^M}, breaklines=true, keywordstyle=\bfseries, morekeywords={type,subtype,break,continue,if,else,end,loop,while,do,done,exit, when,then,return,read,and,or,not,for,each,boolean,procedure,invoke,next,iteration,until,goto}}}

\newcommand{\later}[1]{}

\begin{document}

\title{A Template for Implementing Fast Lock-free Trees Using HTM}

\author{Trevor Brown}
\orcid{0000-0002-0074-1031}
\affiliation{Technion, Israel}
\email{me@tbrown.pro}

\setcopyright{none}
\acmConference{CONF 'YY}{Month DD, 20YY}{City, Country}

\thanks{This work was performed while Trevor Brown was a student at the University of Toronto. Funding was provided by the Natural Sciences and Engineering Research Council of Canada. I would also like thank my supervisor Faith Ellen for her helpful comments on this work, and to Oracle Labs for providing access to the 72-thread Intel machine used in my experiments.}

\clubpenalty=10000
\widowpenalty = 10000

\begin{abstract}
Algorithms that use hardware transactional memory (HTM) must provide a software-only fallback path to guarantee progress.
The design of the fallback path can have a profound impact on performance.
If the fallback path is allowed to run concurrently with hardware transactions, then hardware transactions must be instrumented, adding significant overhead.
Otherwise, hardware transactions must wait for any processes on the fallback path, causing concurrency bottlenecks, or move to the fallback path.
We introduce an approach that combines the best of both worlds.
The key idea is to use three execution paths: an HTM fast path, an HTM middle path, and a software fallback path, such that the middle path can run concurrently with each of the other two.
The fast path and fallback path do \textit{not} run concurrently, so the fast path incurs no instrumentation overhead.
Furthermore, fast path transactions can move to the middle path instead of waiting or moving to the software path.
We demonstrate our approach by producing an accelerated version of the tree update template of Brown et~al., which can be used to implement fast lock-free data structures based on down-trees.
We used the accelerated template to implement two lock-free trees: a binary search tree (BST), and an ($a,b$)-tree (a generalization of a B-tree).
Experiments show that, with 72 concurrent processes, our accelerated ($a,b$)-tree performs between 4.0x and 4.2x as many operations per second as an implementation obtained using the original tree update template.
\end{abstract}

\maketitle

%
%

\section{Introduction}

Concurrent data structures are crucial building blocks in multi-threaded software.
There are many concurrent data structures implemented using locks, but locks can be inefficient, and are not fault tolerant (since a process that crashes while holding a lock can prevent all other processes from making progress).
%
Thus, it is often preferable to use hardware synchronization primitives like compare-and-swap (\textsc{CAS}) instead of locks.
This enables the development of \textit{lock-free} (or \textit{non-blocking}) data structures, which guarantee that at least one process will always continue to make progress, even if some processes crash.
However, it is notoriously difficult to implement lock-free data structures from CAS, and this has inhibited the development of advanced lock-free data structures.

One way of simplifying this task is to use a higher level synchronization primitive that can atomically access multiple locations. 
For example, consider a $k$-word compare-and-swap ($k$-CAS), which atomically: reads $k$ locations, checks if they contain $k$ expected values, and, if so, writes $k$ new values.
$k$-CAS is highly expressive, and it can be used in a straightforward way to implement \textit{any} atomic operation.
Moreover, it can be implemented from CAS and registers~\cite{Harris:2002}.
However, since $k$-CAS is so expressive, it is difficult to implement efficiently. 

Brown et~al.~\cite{Brown:2013} developed a set of new primitives called \llt\ and \sct\ that are less expressive than $k$-CAS, but can still be used in a natural way to implement many advanced data structures.
These primitives can be implemented much more efficiently than $k$-CAS.
At a high level, \llt\ returns a snapshot of a node in a data structure, and after performing \llt s on one or more nodes, one can perform an \sct\ to atomically: change a field of one of these nodes, and \textit{finalize} a subset of them, \textit{only if} none of these nodes have changed since the process performed \llt s on them.
Finalizing a node prevents any further changes to it, which is useful to stop processes from erroneously modifying deleted parts of the data structure.
In a subsequent paper, Brown et~al. used \llt\ and \sct\ to design a \textit{tree update template} that can be followed to produce lock-free implementations of down-trees (trees in which all nodes except the root have in-degree one) with any kinds of update operations~\cite{Brown:2014}.
They demonstrated the use of the template by implementing a chromatic tree, which is an advanced variant of a red-black tree (a type of balanced binary search tree) that offers better scalability. 
The template has also been used to implement many other advanced data structures, including lists, relaxed AVL trees, relaxed ($a,b$)-trees, relaxed $b$-slack trees and weak AVL trees~\cite{Brown:2014,BrownPhD,He:2016}.
Some of these data structures are highly efficient, and would be well suited for inclusion in data structure libraries.

In this work, we study how the new hardware transactional memory (HTM) capabilities found in recent processors (e.g., by Intel and IBM) can be used to produce significantly faster implementations of the tree update template. 
By accelerating 
the tree update template, we also provide a way to accelerate all of the data structures that have been implemented with it.
Since library data structures are reused many times, even minor performance improvements confer a large benefit.

%
HTM allows a programmer to run blocks of code in transactions, which either commit and take effect atomically, or abort and have no effect on shared memory.
Although transactional memory was originally intended to \textit{simplify} concurrent programming, researchers have since realized that HTM can also be used effectively to \textit{improve the performance of existing concurrent code}~\cite{Liu2015,timnat2015practical,makreshanski2015lock}:
Hardware transactions typically have very little overhead, so they can often be used to replace other, more expensive synchronization mechanisms.
For example, instead of performing a sequence of CAS primitives, it may be faster to perform reads, if-statements and writes inside a transaction. 
%
Note that this represents a \textit{non-standard use of HTM}: we are \textit{not} interested in its ease of use, but, rather, in its ability to reduce synchronization costs.

Although hardware transactions are fast, it is surprisingly difficult to obtain the full performance benefit of HTM.
Here, we consider Intel's HTM, which is a \textit{best-effort} implementation.
This means it offers \textit{no guarantee} that transactions will ever commit.
Even in a single threaded system, a transaction can repeatedly abort because of internal buffer overflows, page faults, interrupts, and many other events.
So, to guarantee progress, any code that uses HTM must also provide a \textit{software fallback path} to be executed if a transaction fails.
The design of the fallback path 
profoundly impacts the performance of HTM-based algorithms.

\fakeparagraph{Allowing concurrency between two paths}
Consider an operation $O$ that is implemented using the tree update template.
One natural way to use HTM to accelerate $O$ is to use the original operation as a fallback path, and then obtain an HTM-based fast path by wrapping $O$ in a transaction, and performing optimizations to improve performance~\cite{Liu2015}.
We call this the \textbf{2-path concurrent} algorithm (\textit{2-path con}).
Since the fast path is just an optimized version of the fallback path, transactions on the fast path and fallback path can safely run concurrently.
If a transaction aborts, it can either be retried on the fast path, or be executed on the fallback path.
Unfortunately, supporting concurrency between the fast path and fallback path can add significant overhead on the fast path.

The first source of overhead is \textit{instrumentation} on the fast path that manipulates the \textit{meta-data} used by the fallback path to synchronize processes.
For example, lock-free algorithms often create a \textit{descriptor} for each update operation (so that processes can determine how to help one another make progress), and store pointers to these descriptors in \rec s, where they act as locks.
The fast path must also manipulate these descriptors and pointers so that the fallback path can detect changes made by the fast path.

The second source of overhead comes from constraints imposed by algorithmic assumptions made on the fallback path. 
The tree update template implementation in~\cite{Brown:2014} assumes that only child pointers can change, and all other fields of nodes, such as keys and values, are never changed. 
Changes to these other \textit{immutable} fields must be made by replacing a node with a new copy that reflects the desired change.
%
Because of this assumption on the fallback path, transactions on the fast path \textit{cannot} directly change any field of a node other than its child pointers. 
This is because the fallback path has no mechanism to detect such a change (and may, for example, erroneously delete a node that is concurrently being modified by the fast path).
%
Thus, just like the fallback path, the fast path must replace a node with a new copy to change its immutable fields, which can be much less efficient than changing its fields directly.

\fakeparagraph{Disallowing concurrency between two paths}
To avoid the overheads described above, concurrency is often \textit{disallowed} between the fast path and fallback path.
The simplest example of this approach is a technique called \textbf{transactional lock elision} (TLE) \cite{rajwar2001speculative,rajwar2002transactional}.
TLE is used to implement an operation by wrapping its sequential code in a transaction, and falling back to acquire a global lock after a certain number of transactional attempts.
At the beginning of each transaction, a process reads the state of the global lock and aborts the transaction if the lock is held (to prevent inconsistencies that might arise because the fallback path is not atomic).
Once a process begins executing on the fallback path, all concurrent transactions abort, and processes wait until the fallback path is empty before retrying their transactions.

If transactions never abort, then \textit{TLE represents the best performance we can hope to achieve}, because the fallback path introduces almost no overhead and synchronization is performed entirely by hardware.
Note, however, that TLE is not lock-free.
Additionally, in workloads where operations periodically run on the fallback path, performance can be very poor.

As a toy example, consider a TLE implementation of a binary search tree, with a workload consisting of insertions, deletions and \textit{range queries}.
A range query returns all of the keys in a range [$lo, hi$).
Range queries access many memory locations, and cause frequent transactional aborts due to internal processor buffer overflows (capacity limits).
Thus, range queries periodically run on the fallback path, where they can lead to numerous performance problems.
Since the fallback path is sequential, range queries (or any other long-running operations) cause a severe \textit{concurrency bottleneck}, because they prevent transactions from running on the fast path while they slowly complete, serially.

\begin{figure*}
    \vspace{-4mm}
    \centering
    \begin{minipage}{0.5\textwidth}
        \centering
        \includegraphics[width=\linewidth]{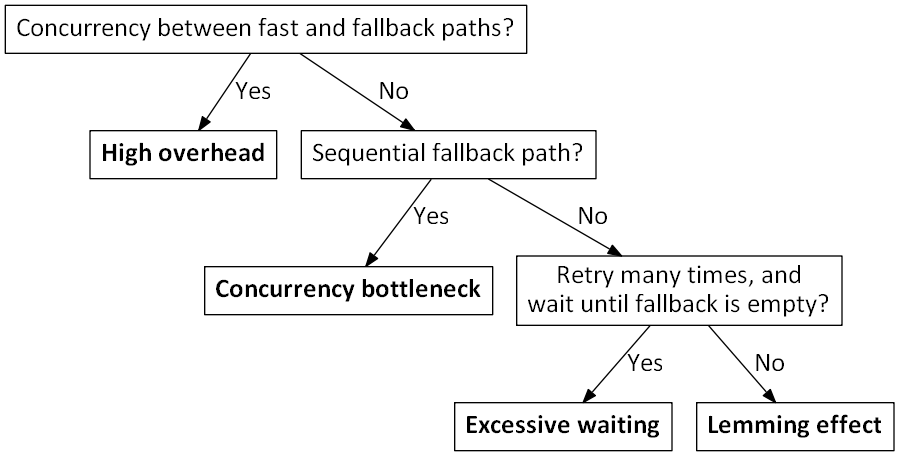}
    \end{minipage}
    \hspace{10mm}
    \begin{minipage}{0.4\textwidth}
        \centering
        \includegraphics[width=\linewidth]{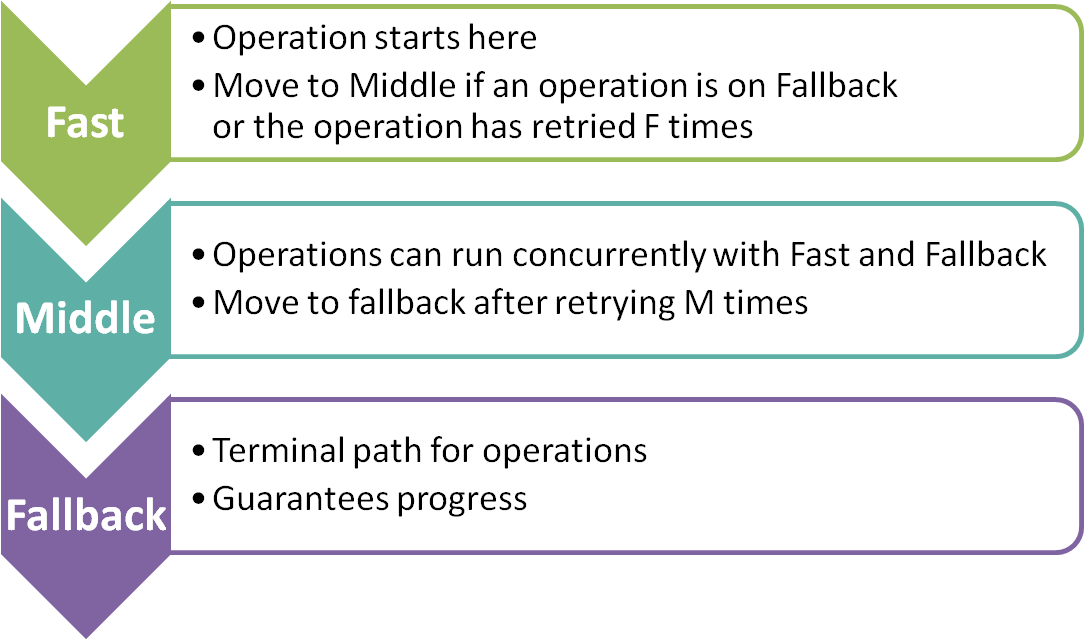}
    \end{minipage}
    \vspace{-3mm}
\caption{(Left) Performance problems affecting two-path algorithms. (Right) Using three execution paths.}
\label{fig-problem}
\end{figure*}

One way to mitigate this bottleneck is to replace the sequential fallback path in TLE with a lock-free algorithm, and replace the global lock with a fetch-and-increment object $F$ that counts how many operations are running on the fallback path. 
Instead of aborting if the lock is held, transactions on the fast path abort if $F$ is non-zero.
We call this the \textbf{2-path non-concurrent} algorithm (\textit{2-path $\overline{con}$}).
In this algorithm, if transactions on the fast path retry only a few times before moving to the fallback path, or do not wait between retries for the fallback path to become empty, then the lemming effect~\cite{Dice2009} can occur. (The lemming effect occurs when processes on the fast path rapidly fail and move to the fallback path, simply because other processes are on the fallback path.)
This can cause the algorithm to run only as fast as the (much slower) fallback path.
However, if transactions avoid the lemming effect by retrying many times before moving to the fallback path, and waiting between retries for the fallback path to become empty, then processes can spend most of their time \textit{waiting}.
The performance problems discussed up to this point are summarized in Figure~\ref{fig-problem}.


\fakeparagraph{The problem with two paths}
In this paper, we study two different types of workloads: \textbf{light workloads}, in which transactions rarely run on the fallback path, and \textbf{heavy workloads}, in which transactions more frequently run on the fallback path.
In light workloads, algorithms that allow concurrency between paths perform very poorly (due to high overhead) in comparison to algorithms that disallow concurrency.
However, in heavy workloads, algorithms that disallow concurrency perform very poorly (since transactions on the fallback path prevent transactions from running on the fast path) in comparison to algorithms that allow concurrency between paths.
Consequently, all two path algorithms have workloads that yield poor performance.
Our experiments confirm this, showing surprisingly poor performance for two path algorithms in many cases.
%

\fakeparagraph{Using three paths}
%
%
We introduce a technique that simultaneously achieves high performance for both light and heavy workloads by using three paths: an HTM fast path, an HTM middle path and a non-transactional fallback path.
(See the right half of Figure~\ref{fig-problem}.)
Each operation begins on the fast path, and moves to the middle path after it retries $F$ times.
An operation on the middle path moves to the fallback path after retrying $M$ times on the middle path.
The fast path does not manipulate any synchronization meta-data used by the fallback path, so operations on the fast path and fallback path cannot run concurrently. 
Thus, whenever an operation is on the fallback path, all operations on the fast path move to the middle path.
The middle path manipulates the synchronization meta-data used by the fallback path, so operations on the middle path and fallback path can run concurrently. 
Operations on the middle path can also run concurrently with operations on the fast path (since conflicts are resolved by the HTM system).
We call this the \textbf{3-path} algorithm (\textit{3-path}).

We briefly discuss why this approach avoids the performance problems described above. 
Since transactions on the fast path do not run concurrently with transactions on the fallback path, transactions on the fast path run with \textit{no instrumentation overhead}.
When a transaction is on the fallback path, transactions can freely execute on the middle path, \textit{without waiting}.
The \textit{lemming effect} does not occur, since transactions do not have to move to the fallback path simply because a transaction is on the fallback path.
Furthermore, we enable a high degree of concurrency, because the fast and middle paths can run concurrently, and the middle and fallback paths can run concurrently.

We performed experiments to evaluate our new template algorithms by comparing them with the original template algorithm.
In order to compare the different template algorithms, we used each algorithm to implement two data structures: a binary search tree (BST) and a relaxed ($a,b$)-tree.
We then ran microbenchmarks to compare the performance (operations per second) of the different implementations in both light and heavy workloads.
The results show that our new template algorithms offer significant performance improvements.
For example, on an Intel system with 72 concurrent processes, our best implementation of the relaxed ($a,b$)-tree outperformed the implementation using the original template algorithm by an average of 410\% over all workloads.

\medskip

\noindent\textbf{Contributions}
\begin{compactitem}
    \item We present four accelerated implementations of the tree update template of Brown~et~al. that explore the design space for HTM-based implementations: \textit{2-path con}, \textit{TLE}, \textit{2-path $\overline{con}$}, and \textit{3-path}.
    \item We highlight the importance of studying both light and heavy workloads in the HTM setting. Each serves a distinct role in evaluating algorithms: light workloads demonstrate the potential of HTM to improve performance by reducing overhead, and heavy workloads capture the performance impact of interactions between different execution paths.
    \item We demonstrate the effectiveness of our approach by accelerating two different lock-free data structures: an unbalanced BST, and a relaxed ($a,b$)-tree.
    Experimental results show a significant performance advantage for our accelerated implementations.
%
\end{compactitem}
\medskip

The remainder of the paper is structured as follows.
The model is introduced in Section~\ref{sec-3path-model}.
Section~\ref{sec-3path-background} describes \llt\ and \sct, and the tree update template.
We describe an HTM-based implementation of \llt\ and \sct\ in Section~\ref{sec-htmscx}.
In Section~\ref{sec-3path-algs}, we describe our four accelerated template implementations, and argue correctness and progress.
In Section~\ref{sec-3path-ds}, we describe two data structures that we use in our experiments.
Experimental results 
are presented in Section~\ref{sec-3path-exp}.
Section~\ref{sec-3path-search-outside} describes an optimization to two of our accelerated template implementations.
In Section~\ref{sec-3path-memrecl}, we describe a way to reclaim memory more efficiently for \textit{3-path} algorithms.
In Section~\ref{sec-3path-other-uses}, we describe how our approach could be used to accelerate data structures that use the read-copy-update (RCU) or $k$-compare-and-swap primitives.
Related work is surveyed in Section~\ref{sec-3path-related}.
Finally, we conclude in Section~\ref{sec-3path-conclusion}.

\section{Model} \label{sec-3path-model}

We consider an asynchronous shared memory system with $n$ processes, and Intel's implementation of HTM.
Arbitrary blocks of code can be executed as transactions, which either commit (and appear to take place instantaneously) or abort (and have no effect on the contents of shared memory).
A transaction is started by invoking \textit{txBegin}, is committed by invoking \textit{txEnd}, and can be aborted by invoking \textit{txAbort}.
Intel's implementation of HTM is best-effort, which means that the system can force transactions to abort at any time, and no transactions are ever guaranteed to commit.
Each time a transaction aborts, the hardware provides a reason why the abort occurred.
Two reasons are of particular interest.
\textit{Conflict} aborts occur when two processes contend on the same cache-line.
Since a cache-line contains multiple machine words, \textit{conflict} aborts can occur even if two processes never contend on the same memory location.
\textit{Capacity} aborts occur when a transaction exhausts some shared resource within the HTM system.
Intuitively, this occurs when a transaction accesses too many memory locations.
(In reality, \textit{capacity} aborts also occur for a variety of complex reasons that make it difficult to predict when they will occur.)

\section{Background} \label{sec-3path-background}

\fakeparagraph{The \llt\ and \sct\ primitives}
The load-link extended (\llt) and store-conditional extended (\sct) primitives are multi-word generalizations of the well-known load-link (LL) and store-conditional (SC), and they have been implemented from single-word \cas\ \cite{Brown:2013}.
\llt\ and \sct\ operate on \rec s, each of which consists of a fixed number of mutable fields (which can change), and a fixed number of immutable fields (which cannot). 

\llt($r$) attempts to take a snapshot of the mutable fields of a \rec\ $r$.
If it is concurrent with an \sct\ involving~$r$, it may return \fail, instead.
Individual fields of a \rec\ can also be read directly.
An \sct($V,$ $R,$ $fld,$ $new$) takes as its arguments a sequence $V$ of \rec s, a subsequence $R$ of $V$, a pointer $fld$ to a mutable field of one \rec\ in~$V$, and a new value $new$ for that field.
The \sct\ tries to atomically store the value $new$ in the field that $fld$ points to and {\it finalize} each \rec\ in $R$.
Once a \rec\ is finalized, its mutable fields cannot be changed by any subsequent \sct, and any \llt\ of the \rec\ will return \finalized\ instead of a snapshot.

Before a process $p$ invokes \sct, it must perform an \llt($r$) on each \rec\ $r$ in $V$.
For each $r \in V$, the last \llt($r$) performed by $p$ prior to the \sct\ is said to be {\it linked} to the \sct, and this linked \llt\ must return a snapshot of $r$ (not \fail\ or \finalized).
An \sct($V, R, fld, new$) by a process modifies the data structure and returns \true\ (in which case we say it \textit{succeeds}) only if no \rec\ $r$ in $V$ has changed since its linked \llt($r$); otherwise the \sct\ fails and returns \false.
Although \llt\ and \sct\ can fail, their failures are limited in such a way that they can be used to build data structures with lock-free progress.
See \cite{Brown:2013} for a more formal specification. 

Observe that \sct\ can only change a single value in a \rec\ (and finalize a sequence of \rec s) atomically.
Thus, to implement an \textit{operation} that changes multiple fields, one must create \textit{new} \rec s that contain the desired changes, and use \sct\ to change \textit{one} pointer to replace the old \rec s. 

%


\begin{figure*}[t]
\small
\def\pwidth{4cm}
\prepnewlisting
\begin{framed}
\vspace{-1.5mm}
\hspace{5mm}
\begin{minipage}{0.57\linewidth}
\def\namewidth{17mm}
\preplisting
\footnotesize
\begin{lstlisting}[mathescape=true,style=nonumbers]
 type $\op$
   //\wcnarrow{$V$}{sequence of \rec s}
   //\wcnarrow{$R$}{subsequence of $V$ to be finalized}
   //\wcnarrow{$fld$}{pointer to a field of a \rec\ in $V$}
   //\wcnarrow{$new$}{value to be written into the field $fld$}
   //\wcnarrow{$old$}{value previously read from the field $fld$} 
   //\wcnarrow{$state$}{one of \{\freezing, \done, \retry\}}
   //\wcnarrow{$\llresults$}{sequence of pointers read from $r.\info$ for each $r \in V$}
   //\wcnarrow{$\freezingdone$}{Boolean}
\end{lstlisting}
\end{minipage}
\begin{minipage}{0.38\linewidth}
\def\namewidth{18mm}
\preplisting
\footnotesize
\begin{lstlisting}[mathescape=true,style=nonumbers]
 type $\rec$
   //\com User-defined fields 
   //\wcnarrow{$m_1, \ldots, m_y$}{mutable fields}
   //\wcnarrow{$i_1, \ldots, i_z$}{immutable fields}
   //\com Fields used by  \llt /\sct\ algorithm
   //\wcnarrow{$\info$}{pointer to an \op}
   //\wcnarrow{$marked$}{Boolean}
   //\vspace{4.5mm}
\end{lstlisting}
\end{minipage}
\vspace{-1.5mm} \hrule \vspace{-1mm}
\footnotesize
\begin{lstlisting}[mathescape=true]
  //\llt$_O(r)$ by process $p$
    $marked_1 := r.marked$ // \label{ll-read-marked1}
    $r\info := r.\info$ // \label{ll-read} 
    $state := r\info.state$ // \label{ll-read-state}
    $marked_2 := r.marked$ // \label{ll-read-marked2}
    if $state = \retry$ or $(state = \done$ and not $marked_2)$ then //  \label{ll-check-frozen} \sidecom{if $r$ was not frozen at line~\ref{ll-read-state}}
      read $r.m_1,...,r.m_y$ //and record the values in local variables $m_1,...,m_y$%
      \label{ll-collect}
      if $r.\info = r\info$ then//\label{ll-reread}\sidecom{if $r.\info$ points to the same} 
        //store $\langle r, r\info, \langle m_1, ..., m_y \rangle \rangle$ in $p$'s local table %
\sidecom{\op\ as on line~\ref{ll-read}}\label{ll-store}
        return $\langle m_1, ..., m_y \rangle$ // \label{ll-return}  \vspace{2mm}
    if ($r\info.state = \done$ or ($r\info.state = \freezing$ and $\help(r\info)))$ and $marked_1$ then// \label{ll-check-finalized}
      return $\finalized$ // \label{ll-return-finalized}
    else
      if $r.\info.state = \freezing$ then $\help(r.\info)$ // \label{ll-help-fail} 
      return $\fail$ // \label{ll-return-fail} \vspace{1mm} \hrule %
\vspace{1mm}
  //\sct$_O(V, R, fld, new)$ by process $p$
  //\tline{\com Preconditions: (\presctlinked) for each $r$ in $V$, $p$ has performed an invocation $I_r$ of \llt$(r)$ linked to this \sct}%
          {\hspace{19.5mm}(\presctabainit) $new$ is not the initial value of $fld$}%
          {\hspace{19.5mm}(\presctaba) for each $r$ in $V$, no $\sct(V', R', fld, new)$ was linearized before $I_r$ was linearized}
    //\dline{Let $\llresults$ be a pointer to a table in $p$'s private memory containing,}%
            {for each $r$ in $V$, the value of $r.\info$ read by $p$'s last \llt$(r)$}
    //Let $old$ be the value for $fld$ returned by $p$'s last \llt$(r)$\vspace{1.5mm}%
    return $\help(\mbox{pointer to new \op} (V, R, fld, new, old, \freezing,  \false, \llresults ))$ // \label{sct-create-op}\label{sct-call-help} \vspace{1mm} \hrule %
\vspace{1mm}
  //\help$(scxPtr)$ 
    //\com \mbox{Freeze all \rec s in $scxPtr.V$ to protect their mutable fields from being changed by other \sct s}
    for each $r$ in $scxPtr.V \mbox{ enumerated in order}$ do//\label{help-fcas-loop-begin}
      //Let $r$\info\ be the pointer indexed by $r$ in $scxPtr.\llresults$ \label{help-rinfo}
      if not $\cas(r.\info,r\info,scxPtr)$ then //\sidecom{\textbf{\fcas}}\label{help-fcas}
        if $r.\info \neq scxPtr$ then // \label{help-check-frozen} 
          //\com \mbox{Could not freeze $r$ because it is frozen for another \sct}
          if $scxPtr.\freezingdone = \true$ then//\sidecom{\textbf{\fcstep}}\label{help-fcstep}
            //\com the \sct\ has already completed successfully 
            return $\true$ // \label{help-return-true-loop} 
          else
            //\com Atomically unfreeze all \rec s frozen for this \sct 
            $scxPtr.state := \retry$ //\sidecom{\textbf{\astep}}\label{help-astep}
            return $\false$ // \label{help-return-false} \vspace{2mm}
    //\com Finished freezing \rec s (Assert: $state \in \{\freezing, \done\}$) 
    $scxPtr.\freezingdone := \true$//\sidecom{\textbf{\fstep}}\label{help-fstep}
    for each $r \in scxPtr.R$ do $r.marked := \true$ //\sidecom{\textbf{\markstep}}\label{help-markstep}
    //$\cas(scxPtr.fld, scxPtr.old, scxPtr.new)$ \sidecom{\textbf{\upcas}}\label{help-upcas} \vspace{2mm}
    //\com Finalize all $r$ in $R$, and unfreeze all $r$ in $V$ that are not in $R$ 
    $scxPtr.state := \done$//\sidecom{\textbf{\cstep}}\label{help-cstep}
    return $\true$ // \label{help-return-true}
\end{lstlisting}
\vspace{-1mm}
\end{framed}
    \vspace{-4mm}
	\caption{Data types and pseudocode for the original \llt\ and \sct\ algorithm.}
	\label{code-3path-scxo}
\end{figure*}

%
Pseudocode for the original, CAS-based implementation of \llt\ and \sct\ appears in Figure~\ref{code-3path-scxo}.
Each invocation $S$ of \sct$_O(V,$ $R,$ $fld,$ $new)$ starts by creating an \op\ $D$, which contains all of the information necessary to perform $S$, and then invokes \func{Help}$(D)$ to perform it.
The \op\ also contains a $state$ field, which initially contains the value \freezing.
When $S$ finishes, the $state$ field of $D$ will contain either \done\ or \retry\, depending on whether the $S$ succeeded.

$S$ synchronizes with other invocations of \sct$_O$ by taking a special kind of lock on each \rec\ in $V$.
These locks grant exclusive access to an \textit{operation}, rather than to a \textit{process}.
Henceforth, we use the term \textit{freezing} (resp., unfreezing), instead of locking (resp., unlocking), to differentiate this kind of locking from typical mutual exclusion.
A \rec\ $u$ is \textit{frozen} for $S$ if $u.\info$ points to $D$, and either $D.state =$ \done\ and $u.marked = true$ (in which case we say $u$ is \textit{finalized}), or $D.state = \freezing$.

So, $S$ freezes a \rec\ $u$ by using CAS to store a pointer to $D$ in $u.\info$ (at the \fcas\ step in Figure~\ref{code-3path-scxo}).
Suppose $S$ successfully freezes all \rec s in its $V$ sequence.
Then, $S$ prepares to finalize each \rec\ $u \in R$ by setting a \textit{marked} bit in $u$ (at the \markstep\ in Figure~\ref{code-3path-scxo}).
Finally, $S$ changes $fld$ to $new$, and atomically releases all locks by setting $D.state$ to \done\ (at the \cstep\ in Figure~\ref{code-3path-scxo}). 
Observe that setting $D.state$ to \done\ has the effect of atomically finalizing all \rec s in $R$ and unfreezing all \rec s in $V \setminus R$.

Now, suppose $S$ was prevented from freezing some \rec\ $u$ because another invocation $S'$ of \sct$_O$ had already frozen $u$ (i.e., the \fcas\ step by $S$ failed, and it saw $r.\info \neq scxPtr$ at the following line).
Then, $S$ aborts by setting the $D.state$ to \retry\ (at the \astep\ in Figure~\ref{code-3path-scxo}).
This has the effect of atomically unfreezing any \rec s $S$ had frozen.
Note that, before the process that performed $S$ can perform another invocation of \sct$_O$ with $u$ in its $V$-sequence, it must perform \llt$(u)$.
If $u$ is still frozen for $S'$ when this \llt$(u)$ is performed, then the \llt\ will use the information stored in the \op\ at $u$ to \textit{help} $S'$ complete and unfreeze $u$.
(\op s also contain another field $\freezingdone$ that is used to coordinate any processes helping the \sct, ensuring that they do not make conflicting changes to the $state$ field.)

The correctness argument is subtle, and we leave the details to~\cite{Brown:2013}, but one crucial algorithmic property is relevant to our work:

\begin{compactenum}[\hspace{3.4mm}{\bf P1}:]
\item Between any two changes to (the user-defined fields of) a \rec\ $u$, a pointer to a new \op\ (that has never before been contained in $u.\info$) is stored in $u.\info$.
\end{compactenum}

\noindent
This property is used to determine whether a \rec\ has changed between the last \llt$_O$ on it and a subsequent invocation of \sct$_O$.
Consider an invocation $S$ of \sct$_O$($V,R,fld,new$) by a process $p$.
Let $u$ be any \rec\ in $V$, and $L$ be the last invocation of \llt$_O(u)$ by $p$.
$L$ reads $u.\info$ and sees some value $ptr$.
$S$ subsequently performs a \fcas\ step to change $u.\info$ from $ptr$ to point to its \op, freezing $u$.
If this CAS succeeds, then $S$ infers that $u$ has not changed between the read of $u.\info$ in the \llt\ and the \fcas\ step.


\fakeparagraph{Progress properties}
Specifying a progress guarantee for \llt\ and \sct\ operations is subtle, because if processes repeatedly perform \llt\ on \rec s that have been finalized, or repeatedly perform failed \llt s, then they may never be able to invoke \sct.
In particular, it is not sufficient to simply prove that \llt s return snapshots infinitely often, since \textit{all} of the \llt s in a sequence must return snapshots before a process can invoke \sct.
To simplify the progress guarantee for \llt\ and \sct, we make a definition.
An \sct-\func{Update} algorithm is one that performs \llt s on a sequence $V$ of \rec s and invokes \sct($V, R, fld, new$) if they all return snapshots.
The progress guarantee in~\cite{Brown:2013} is then stated as follows.

\smallskip
\begin{compactenum}[\hspace{3.4mm}{\bf PROG}:]
\item Suppose that
    (a) there is always some non-finalized \rec\ reachable by following pointers from an entry point, 
    (b) for each \rec\ $r$, each process performs finitely many invocations of \llt$(r)$ that return \finalized, and
    (c) processes perform infinitely many executions of \sct-\func{Update} algorithms.
    Then, infinitely many invocations of \sct\ succeed.
\end{compactenum}

\begin{figure}
    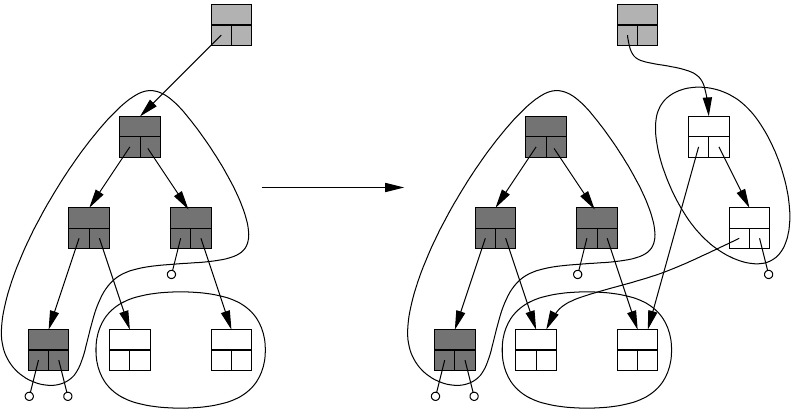
    \caption{Example of the tree update template.
    }
    \label{fig-3path-replace-subtree}
\end{figure}

\fakeparagraph{The tree update template}
The tree update template implements lock-free updates that atomically replace an old connected subgraph $R$ of a down-tree by a new connected subgraph $N$ (as shown in Figure~\ref{fig-3path-replace-subtree}).
Such an update can implement any change to the tree, such as an insertion into a BST or a rotation in a balanced tree.
The old subgraph includes all nodes with a field to be modified.
The new subgraph may have pointers to nodes in the old tree.
Since every node in a down-tree has indegree one, the update can be performed by changing a single child pointer of some node $parent$.
However, problems could arise if a concurrent operation changes the part of the tree being updated.
For example, nodes in the old subgraph, or even $parent$, could be removed from the tree before $parent$'s child pointer is changed.
The template takes care of the process coordination required to prevent such problems.


Each tree node is represented by a \rec\ with a fixed number of child pointers as its mutable fields. 
Each child pointer either points to a \rec\ or contains \nil\ (denoted by $\multimap$ in our figures).
Any other data in the node is stored in immutable fields.
Thus, if an update must change some of this data, it makes a new copy of the node with the updated data.
There is a \rec\ $entry$ which acts as the entry point to the data structure and is never deleted.

At a high level, an update that follows the template proceeds in two phases: the \textit{search phase} and the \textit{update phase}.
In the search phase, the update searches for a location where it should occur.
Then, in the update phase, the update performs \llt s on a connected subgraph of nodes in the tree, including $parent$ and the set $R$ of nodes to be removed from the tree.
Next, it decides whether the tree should be modified, and, if so, creates a new subgraph of nodes and performs an \sct\ that atomically changes a child pointer, as shown in Figure~\ref{fig-3path-replace-subtree}, and finalizes any nodes in $R$.
See~\cite{Brown:2014} for further details.

\section{HTM-based LLX and SCX} \label{sec-htmscx}

In this section, we describe an HTM-based implementation of \llt\ and \sct.
This implementation is used by our first accelerated template implementation, \textit{2-path con}, which is described in Section~\ref{sec-3path-algs}.

\begin{figure*}[th]
\small
\def\pwidth{4cm}
\begin{minipage}{0.48\linewidth}
\begin{framed}
\prepnewlisting
\begin{lstlisting}[mathescape=true]
 //Private variable for process $p$: $\textit{attempts}_p, \textit{tagseq}_p$ \vspace{1mm}\hrule\vspace{1mm}
 //\sct($V,$ $R,$ $fld,$ $new$) by process $p$
 //onAbort: \hfill\com jump here on transaction abort
   if $\mbox{we jumped here after an \textbf{explicit} abort}$ then return $\false$ //\label{newscx-scx-explicit-return}
   if $\textit{attempts}_p < \textit{AttemptLimit}$ then
     $\textit{attempts}_p := \textit{attempts}_p + 1$
     //$retval := \sct_{HTM}$($V,R,fld,new$) \com Fast
   else
     //$retval := \sct_O$($V,R,fld,new$) \com Fallback \vspace{1.5mm}
   if $retval$ then $\textit{attempts}_p := 0$
   return $retval$ //\vspace{-1mm}
\end{lstlisting}
\end{framed}
\end{minipage}
\hspace{0.03\linewidth}
\begin{minipage}{0.48\linewidth}
\begin{framed}
\preplisting
\begin{lstlisting}[mathescape=true]
 //\sct$_{HTM}$($V, R, fld, new$) by process $p$
   //\dline{Let $\llresults$ be a pointer to a table in $p$'s private memory containing,}%
           {for each $r$ in $V$, the value of $r.\info$ read by $p$'s last \llt($r$)}
   //Let $old$ be the value for $fld$ returned by $p$'s last \llt($r$)\vspace{1.5mm}%

   //Begin hardware transaction \label{newscx-scxhtm-xbegin}
   //\textit{tagseq}$_p$ := \textit{tagseq}$_p$ + $2^{\lceil \log n \rceil}$
   for each $r \in V$ do //\label{newscx-scxhtm-freezing-loop-start}
     //Let $r$\info\ be the pointer indexed by $r$ in $\llresults$
     if $r.\info \neq r\info$ then //Abort hardware transaction (explicitly) \label{newscx-scxhtm-abort} \label{newscx-scxhtm-freezing-loop-end}
   for each $r \in V$ do $r.\info := \textit{tagseq}_p$
   for each $r \in R$ do $r.marked := \true$
   //write $new$ to the field pointed to by $fld$
   //Commit hardware transaction \label{newscx-scxhtm-commit}
   return $\true$ //\vspace{-1mm}
\end{lstlisting}
\end{framed}
\end{minipage}
	\caption{Final HTM-based implementation of \sct.}
	\label{code-3path-newscx}
\end{figure*}

In the following, we use \sct$_O$ and \llt$_O$ to refer to the original lock-free implementation of \llt\ and \sct.
We give an implementation of \sct\ that uses an HTM-based fast path called \sct$_{HTM}$, and \sct$_O$ as its fallback path.
Hardware transactions are instrumented so they can run concurrently with processes executing \sct$_O$.
This algorithm guarantees lock-freedom and achieves a high degree of concurrency.
Pseudocode appears in Figure~\ref{code-3path-newscx}.
%
At a high level, an \sct$_{HTM}$ by a process $p$ starts a transaction, then attempts to perform a highly optimized version of \sct$_O$.
Each time a transaction executed by $p$ aborts, control jumps to the onAbort label, at the beginning of the \sct\ procedure.
If a process \textit{explicitly} aborts a transaction at line~\ref{newscx-scxhtm-abort}, 
then \sct\ returns \false\ at line~\ref{newscx-scx-explicit-return}.
Each process has a budget \textit{AttemptLimit} that specifies how many times it will attempt hardware transactions before it will fall back to executing \sct$_O$.

In \sct$_O$, \op s are used (1) to facilitate helping, and (2) to lock \rec s and detect changes to them.
In particular, \sct$_O$ guarantees the following property.
\textbf{P1}: between any two changes to (the user-defined fields of) a \rec\ $u$, a new \op\ pointer is stored in $u.\info$.
However, \sct$_{HTM}$ does not create \op s.
In a transactional setting, helping causes unnecessary aborts, since executing a transaction that performs the same work as a running transaction will cause at least one (and probably both) to abort.
Helping in transactions is also \textit{not necessary} to guarantee progress, since progress is guaranteed by the fallback path.
So, to preserve property P1, we give each process $p$ a \textit{tagged sequence number} \textit{tseq}$_p$ that contains the process name, a sequence number, and a \textit{tag} bit.
The \textit{tag} bit is the least significant bit.
On modern systems where pointers are word aligned, the least significant bit in a pointer is always zero.
Thus, the tag bit allows a process to distinguish between a tagged sequence number and a pointer.
In \sct$_{HTM}$, instead of having $p$ create a new \op\ and store pointers to it in \rec s to lock them, $p$ increments its sequence number in $\textit{tseq}_p$ and stores $\textit{tseq}_p$ in \rec s.
Since no writes performed by a transaction $T$ can be seen until it commits, it never actually needs to hold any locks.
Thus, every value of $\textit{tseq}_p$ stored in a \rec\ represents an unlocked value, and writing $\textit{tseq}_p$ represents $p$ locking and immediately unlocking a node.

After storing $\textit{tseq}_p$ in each $r \in V$, \sct$_{HTM}$ finalizes each $r \in R$ by setting $r.marked := \true$ (mimicking the behaviour of \sct$_O$).
Then, it stores $new$ in the field pointed to by $fld$, and commits.
Note that eliminating the creation of \op s on the fast path also eliminates the need to \textit{reclaim} any created \op s, which further reduces overhead.

The \sct$_{HTM}$ algorithm also necessitates a small change to \llt$_O$, to handle tagged sequence numbers.
An invocation of \llt$_O$($r$) reads a pointer $r\info$ to an \op, follows $r\info$ to read one of its fields, and uses the value it reads to determine whether $r$ is locked.
However, $r\info$ may now contains a tagged sequence number, instead of a pointer to an \op.
So, in our modified algorithm, which we call \llt$_{HTM}$, before a process tries to follow $r\info$, it first checks whether $r\info$ is a tagged sequence number, and, if so, behaves as if $r$ is unlocked.
The code for \llt$_{HTM}$ appears in Figure~\ref{code-3path-htmsct-transformation3}.

\subsection{Correctness and Progress}

The high-level idea is to show that one can start with \llt$_O$ and \sct$_O$, and obtain our HTM-based implementation by applying a sequence of transformations.
Intuitively, these transformations preserve the semantics of \sct\ and maintain backwards compatibility with \sct$_O$ so that the transformed versions can be run concurrently with invocations of \sct$_O$.
More formally, for each execution of a transformed algorithm, there is an execution of the original algorithm in which: the same operations are performed, they are linearized in the same order, and they return the same results.
For each transformation, we sketch the correctness and progress argument, since the transformations are simple and a formal proof would be overly pedantic.

\fakeparagraph{Adding transactions}
\begin{figure*}[th]
\footnotesize
\def\pwidth{4cm}
\prepnewlisting
\vspace{-2mm}
\begin{framed}
\begin{lstlisting}[mathescape=true]
  //\sct$_1$($V, R, fld, new$) by process $p$
    //\dline{Let $\llresults$ be a pointer to a table in $p$'s private memory containing,}%
            {for each $r$ in $V$, the value of $r.\info$ read by $p$'s last \llt($r$)}
    //Let $old$ be the value for $fld$ returned by $p$'s last \llt($r$)\vspace{1.5mm}%

    //Begin hardware transaction
    $scxPtr := \mbox{pointer to new \op}(V,R,fld,new,old,\freezing,\false,\llresults$) //\label{htmsct-transformation1-new-op}
    //\com \mbox{Freeze all \rec s in $scxPtr.V$ to protect their mutable fields from being changed by other \sct s}
    for each $r$ in $scxPtr.V \mbox{ enumerated in order}$ do//\label{htmsct-transformation1-fcas-loop-begin}
      //Let $r$\info\ be the pointer indexed by $r$ in $scxPtr.\llresults$ \label{htmsct-transformation1-rinfo}
      if not $\cas$($r.\info,r\info,scxPtr$) then //\sidecom{\textbf{\fcas}}\label{htmsct-transformation1-fcas}
        if $r.\info \neq scxPtr$ then // \label{htmsct-transformation1-check-frozen} 
          //\com \mbox{Could not freeze $r$ because it is frozen for another \sct}
          if $scxPtr.\freezingdone = \true$ then//\sidecom{\textbf{\fcstep}}\label{htmsct-transformation1-fcstep}
            //\com the \sct\ has already completed successfully 
            //Commit hardware transaction\label{htmsct-transformation1-commit1}
            return $\true$ // \label{htmsct-transformation1-return-true-loop} 
          else //Abort hardware transaction (explicitly) \label{htmsct-transformation1-commit2} \vspace{1.5mm}%

    //\com Finished freezing \rec s (Assert: $state \in \{\freezing, \done\}$) 
    $scxPtr.\freezingdone := \true$//\sidecom{\textbf{\fstep}}\label{htmsct-transformation1-fstep}
    for each $r \in scxPtr.R$ do $r.marked := \true$ //\sidecom{\textbf{\markstep}}\label{htmsct-transformation1-markstep}
    //$\cas$($scxPtr.fld, scxPtr.old, scxPtr.new$) \sidecom{\textbf{\upcas}}\label{htmsct-transformation1-upcas} \vspace{2mm}
    //\com Finalize all $r$ in $R$, and unfreeze all $r$ in $V$ that are not in $R$ 
    $scxPtr.state := \done$//\sidecom{\textbf{\cstep}}\label{htmsct-transformation1-cstep}
    //Commit hardware transaction\label{htmsct-transformation1-commit3}
    return $\true$ // \label{htmsct-transformation1-return-true} \vspace{-1mm}
\end{lstlisting}
\end{framed}
    \vspace{-5mm}
	\caption{Transforming \sct$_O$: after adding transactions.}
	\label{code-3path-htmsct-transformation1}
\end{figure*}
For the first transformation, we replaced the invocation of \help\ in \sct$_O$ with the body of the \help\ function, and wrapped the code in a transaction.
Since the fast path simply executes the fallback path algorithm in a transaction, the correctness of the resulting algorithm is immediate from the correctness of the original \llt\ and \sct\ algorithm.

We also observe that it is not necessary to commit a transaction that sets the $state$ of its \op\ to \retry\ and returns \false.
The only effect that committing such a transaction would have on shared memory is changing some of the $\info$ fields of \rec s in its $V$ sequence to point to its \op.
In \sct$_O$, $\info$ fields serve two purposes.
First, they provide pointers to an \op\ while its \sct\ is in progress (so it can be helped).
Second, they act as locks that grant exclusive access to an \sct$_O$, and allow an invocation of \sct$_O$ to determine whether any user-defined fields of a \rec\ $r$ have changed since its linked \llt($r$) (using property P1).
However, since the effects of a transaction are not visible until it has already committed, a 
transaction no longer needs help by the time it modified any $\info$ field.
And, since an \sct$_O$ that sets the $state$ of its \op\ to \retry\ does not change any user-defined field of a \rec, these changes to $\info$ fields are not needed to preserve property P1.
The only consequence of changing these $\info$ fields is that other invocations of \sct$_O$ might needlessly fail and return \false, as well.
So, instead of setting $state = \retry$ and committing, we \textit{explicitly abort} the transaction and return \false.
Figure~\ref{code-3path-htmsct-transformation1} shows the result of this transformation: \sct$_1$.
(Note that aborting transactions does not affect correctness---only progress.)

\begin{figure*}[th]
\footnotesize
\def\pwidth{4cm}
\prepnewlisting
\vspace{-2mm}
\begin{framed}
\begin{lstlisting}[mathescape=true]
  //Private variable for process $p$: $\textit{attempts}_p$\vspace{1mm}\hrule\vspace{1mm}
  //\sct($V, R, fld, new$) by process $p$
  //onAbort: \hfill\com jump here on transaction abort \label{htmsct-usage-onabort}
    if $\mbox{we jumped here after an explicit abort in the code}$ then return $\false$
    if $\textit{attempts}_p < \textit{AttemptLimit}$ then //\label{htmsct-usage-check-attempts}
      $\textit{attempts}_p := \textit{attempts}_p + 1$
      $retval := \sct_1$($V,R,fld,new$) //\hfill\com invoke HTM-based \sct
    else
      $retval := \sct_O$($V,R,fld,new$) //\hfill\com fall back to original \sct\vspace{1.5mm}
    if $retval$ then $\textit{attempts}_p := 0$//\hfill\com reset $p$'s attempt counter before returning \true\label{htmsct-usage-return-true}
    return $retval$ //\vspace{-1mm}
\end{lstlisting}
\end{framed}
    \vspace{-4mm}
	\caption{How the HTM-based \sct$_1$ is used to provide lock-free \sct.}
	\label{code-3path-htmsct-usage}
\end{figure*}

Of course, we must provide a fallback code path in order to guarantee progress.
Figure~\ref{code-3path-htmsct-usage} shows how \sct$_1$ (the fast path) and \sct$_O$ (the fallback path) are used together to implement lock-free \sct.
In order to decide when each code path should be executed, we give each process $p$ a private variable $\textit{attempts}_p$ that contains the number of times $p$ has attempted a hardware transaction since it last performed an \sct$_1$ or \sct$_O$ that succeeded (i.e., returned \true).
The \sct\ procedure checks whether $\textit{attempts}_p$ is less than a (positive) threshold \textit{AttemptLimit}.
If so, $p$ increments $\textit{attempts}_p$ and invokes \sct$_1$ to execute a transaction on the fast path.
If not, $p$ invokes \sct$_O$ (to guarantee progress).
Whenever $p$ returns \true\ from an invocation of \sct$_1$ or \sct$_O$, it resets its budget $\textit{attempts}_p$ to zero, so it will execute on the fast path in its next \sct.
Each time a transaction executed by $p$ aborts, control jumps to the onAbort label, at the beginning of the \sct\ procedure.
If a process explicitly aborts a transaction it is executing (at line~\ref{htmsct-transformation1-commit2} in $\sct_1$), 
then control jumps to the onAbort label, \textit{and} the \sct\ returns \false\ at the next line.

\fakeparagraph{Progress}
%
It is proved in~\cite{Brown:2013} that PROG is satisfied by \llt$_O$ and \sct$_O$.
We argue that PROG is satisfied by the implementation of \llt\ and \sct\ in Figure~\ref{code-3path-htmsct-usage}. 
To obtain a contradiction, suppose the antecedent of PROG holds, but only finitely many invocations of \sct\ return \true.
Then, after some time $t$, no invocation of \sct\ returns \true.

\textit{Case 1:} Suppose processes take infinitely many steps in transactions.
By inspection of the code, each transaction is wait-free, and \sct\ returns \true\ immediately after a transaction commits.
Since no transaction commits after $t$, there must be infinitely many aborts.
However, each process can perform at most \textit{AttemptLimit} aborts since the last time it performed an invocation of \sct\ that returned \true.
So, only finitely many aborts can occur after $t$---a contradiction.

\textit{Case 2:} Suppose processes take only finitely many steps in transactions.
Then, processes take only finitely many steps in \sct$_1$.
It follows that, after some time $t'$, no process takes a step in \sct$_1$.
Therefore, in the suffix of the execution after $t'$, processes only take steps in \sct$_O$ and \llt$_O$.
However, since \llt$_O$ and \sct$_O$ satisfy PROG, infinitely many invocations of \sct\ must succeed after $t'$, which is a contradiction.

\fakeparagraph{Eliminating \textit{most} accesses to fields of \op s created on the fast path}
\begin{figure*}[th]
\footnotesize
\def\pwidth{4cm}
\prepnewlisting
\vspace{-2mm}
\begin{framed}
\begin{lstlisting}[mathescape=true]
  //Private variable for process $p$: $\textit{attempts}_p$\vspace{1mm}\hrule\vspace{1mm}
  //\sct$_2$($V, R, fld, new$) by process $p$
    //\dline{Let $\llresults$ be a pointer to a table in $p$'s private memory containing,}%
            {for each $r$ in $V$, the value of $r.\info$ read by $p$'s last \llt($r$)}
    //Let $old$ be the value for $fld$ returned by $p$'s last \llt($r$)\vspace{1.5mm}%

    //Begin hardware transaction
    $scxPtr := \mbox{pointer to new \op}$($-,-,-,-,-,\freezing,-,-$) //\label{htmsct-transformation2-new-op}
    //\com \mbox{Freeze all \rec s in $V$ to protect their mutable fields from being changed by other \sct s}
    for each $r$ in $V \mbox{ enumerated in order}$ do//\label{htmsct-transformation2-fcas-loop-begin}
      //Let $r$\info\ be the pointer indexed by $r$ in $\llresults$ \label{htmsct-transformation2-rinfo}
      if not $\cas$($r.\info,r\info,scxPtr$) then //\sidecom{\textbf{\fcas}}\label{htmsct-transformation2-fcas}
        if $r.\info \neq scxPtr$ then //Abort hardware transaction (explicitly)\vspace{1.5mm}%

    //\com Finished freezing \rec s
    for each $r \in R$ do $r.marked := \true$ //\hspace{4mm}\com Finalize each $r \in R$ \sidecom{\textbf{\markstep}}\label{htmsct-transformation2-markstep}
    //$\cas$($fld, old, new$) \sidecom{\textbf{\upcas}}\label{htmsct-transformation2-upcas}
    $scxPtr.state := \done$//\sidecom{\textbf{\cstep}}\label{htmsct-transformation2-cstep}
    //Commit hardware transaction
    return $\true$ // \label{htmsct-transformation2-return-true}\vspace{-1mm}
\end{lstlisting}
\end{framed}
    \vspace{-4mm}
	\caption{Transforming \sct$_O$: after eliminating \textbf{most} accesses to fields of \op s created on the fast path.}
	\label{code-3path-htmsct-transformation2}
\end{figure*}
In \llt$_O$ and \sct$_O$, helping is needed to guarantee progress, because otherwise, 
an invocation of \sct$_O$ that crashes while one or more \rec s are frozen for it could cause every invocation of \llt$_O$ to return \fail\ (which, in turn, could prevent processes from performing the necessary linked invocations of \llt$_O$ to invoke \sct$_O$).
However, as we mentioned above, since transactions are atomic, a process cannot see any of their writes (including the contents of any \op\ they create and publish pointers to) until they have committed, at which point they no longer need help.
Thus, it is not necessary to help transactions in \sct$_1$.\footnote{In fact, helping transactions would be \textit{actively harmful}, since performing the same modifications to shared memory as an in-flight transaction \textit{will cause it to abort}. This leads to very poor performance, in practice.}

In fact, it is easy to see that processes will not help any \op\ created by a transaction in \sct$_1$.
Observe that each transaction in $\sct_1$ sets the $state$ of its \op\ to \done\ before committing.
Consequently, if an invocation of \llt$_O$ reads $r.\info$ and obtains a pointer $r\info$ to an \op\ created by a transaction in $\sct_1$, then $r\info$ has $state$ \done.
Therefore, by inspection of the code, \llt$_O$ will not invoke \help($r\info$).

Since \llt$_O$ never invokes \help($r\info$) for any $r\info$ created by a transaction in \sct$_1$, most fields of an \op\ created by a transaction are accessed only by the process that created the \op.
The only field that is accessed by other processes is the $state$ field (which is accessed in \llt$_O$).
Therefore, it suffices for a transaction in \sct$_1$ to initialize only the $state$ field of its \op.
As we will see, any accesses to the other fields can simply be eliminated or replaced with locally available information.

Using this knowledge, we transform $\sct_1$ in Figure~\ref{code-3path-htmsct-transformation1} into a new procedure called $\sct_2$ in Figure~\ref{code-3path-htmsct-transformation2}.
First, instead of initializing the entire \op\ when we create a new \op\ at line~\ref{htmsct-transformation1-new-op} in $\sct_1$, we initialize only the $state$ field.
We then change any steps that read fields of the \op\ (lines~\ref{htmsct-transformation1-fcas-loop-begin}, \ref{htmsct-transformation1-rinfo}, \ref{htmsct-transformation1-fcstep}, \ref{htmsct-transformation1-markstep} and~\ref{htmsct-transformation1-upcas} in $\sct_1$) to use locally available information, instead.

Next, we eliminate the \textit{\fstep} at line~\ref{htmsct-transformation1-fstep} in $\sct_1$, which changes the $\freezingdone$ field of the \op.
Recall that $\freezingdone$ is used by \sct$_O$ to prevent helpers from making conflicting changes to the $state$ field of its \op.
When a \textit{\fcas} fails in an invocation $S$ of \sct$_O$ (at line~\ref{help-fcas} of \func{Help} in Figure~\ref{code-3path-scxo}), it indicates that either $S$ will fail due to contention, or another process had already helped $S$ to complete successfully.
The $\freezingdone$ bit allows a process to distinguish between these two cases.
Specifically, it is proved in~\cite{Brown:2013} that a process will see $\freezingdone = \true$ at line~\ref{help-fcas} of \func{Help} if and only if another process already helped $S$ complete and set $\freezingdone := \true$.
However, since we have argued that processes never help transactions (and, in fact, no other process can even \textit{access} the \op\ until the transaction that created it has committed), $\freezingdone$ is always \false\ at the corresponding step (line~\ref{htmsct-transformation1-fcstep}) in $\sct_1$.
This observation allows us to eliminate the entire \textit{if} branch at line~\ref{htmsct-transformation1-fcstep} in \sct$_1$.

Clearly, this transformation preserves PROG.
%
Note that \sct$_2$ (and each of the subsequent transformed variants) is used in the same way as \sct$_1$: Simply replace \sct$_1$ in Figure~\ref{code-3path-htmsct-usage} with \sct$_2$.

\fakeparagraph{Completely eliminating accesses to fields of \op s created on the fast path}
\begin{figure*}[th]
\footnotesize
\def\pwidth{4cm}
\prepnewlisting
\vspace{-2mm}
\begin{framed}
\begin{lstlisting}[mathescape=true]
  //Private variable for process $p$: $\textit{attempts}_p$ \vspace{1mm}\hrule\vspace{1mm}
  //\llt$_{HTM}$($r$) by process $p$ \com Precondition: $r \neq \nil$.
    $marked_1 := r.marked$ // \hfill\com{order of lines~\ref{ll-read-marked1}--\ref{ll-read-marked2} matters} \label{htmllt-read-marked1}
    $r\info := r.\info$ // \label{htmllt-read} 
 *  $state := (r\info\ \&\ 1)\ ?\ \mbox{\done}\ : r\info.state$ //\hfill\com if \textit{rinfo} is tagged, take \textit{state} to be \done\ \label{htmllt-read-state}
    $marked_2 := r.marked$ // \label{htmllt-read-marked2}
    if $state = \retry$ or ($state = \done$ and not $marked_2$) then //\hfill\com{if $r$ was not frozen at line~\ref{ll-read-state}} \label{htmllt-check-frozen}
      read $r.m_1,...,r.m_y$ //and record the values in local variables $m_1,...,m_y$%
      \label{htmllt-collect}
      if $r.\info = r\info$ then//\hfill\com{if $r.\info$ points to the same \op\ as on line~\ref{ll-read}} \label{htmllt-reread}
        //store $\langle r, r\info, \langle m_1, ..., m_y \rangle \rangle$ in $p$'s local table \label{htmllt-store}
        return $\langle m_1, ..., m_y \rangle$ // \label{htmllt-return}  \vspace{2mm}
 *  $state_2 := (r\info\ \&\ 1)\ ?\ \mbox{\done}\ : r\info.state$ //\hfill\com if $r\info$ is tagged, take $state_2$ to be \done\ \label{htmllt-state2}
 *  if ($state_2 = \done$ or ($state_2 = \freezing$ and $\help$($r\info$))) and $marked_1$ then//%
    \label{htmllt-check-finalized}
      return $\finalized$ // \label{htmllt-return-finalized}
    else
 *    $r\info_2 := r.\info$
 *    $state_3 := (r\info_2\ \&\ 1)\ ?\ \mbox{\done}\ : r\info_2.state$ //\hfill\com if $r\info_2$ is tagged, take $state_3$ to be \done\ \label{htmllt-state3}
 *    if $state_3 = \freezing$ then $\help$($r\info_2$) // \label{htmllt-help-fail} 
      return $\fail$ // \label{htmllt-return-fail} \vspace{1.5mm} \hrule \vspace{1.5mm}%

  //\sct$_3$($V, R, fld, new$) by process $p$
    //\dline{Let $\llresults$ be a pointer to a table in $p$'s private memory containing,}%
            {for each $r$ in $V$, the value of $r.\info$ read by $p$'s last \llt($r$)}
    //Let $old$ be the value for $fld$ returned by $p$'s last \llt($r$)\vspace{1.5mm}%

    //Begin hardware transaction
    $scxPtr := \mbox{pointer to new \op}$($-,-,-,-,-,-,-,-$) //\label{htmsct-transformation3-new-op}
    //\com \mbox{Freeze all \rec s in $V$ to protect their mutable fields from being changed by other \sct s}
    for each $r$ in $V \mbox{ enumerated in order}$ do//\label{htmsct-transformation3-fcas-loop-begin}
      //Let $r$\info\ be the pointer indexed by $r$ in $\llresults$ \label{htmsct-transformation3-rinfo}
      if not $\cas$($r.\info,r\info, (scxPtr\ \&\ 1)$) then //\sidecom{\textbf{\fcas}}\label{htmsct-transformation3-fcas}
        if $r.\info \neq (scxPtr\ \&\ 1$) then //Abort hardware transaction (explicitly)\label{htmsct-transformation3-abort}\vspace{1.5mm}%

    //\com Finished freezing \rec s
    for each $r \in R$ do $r.marked := \true$ //\hspace{4mm}\com Finalize each $r \in R$ \sidecom{\textbf{\markstep}}\label{htmsct-transformation3-markstep}
    //$\cas$($fld, old, new$) \sidecom{\textbf{\upcas}}\label{htmsct-transformation3-upcas}
    //Commit hardware transaction\label{htmsct-transformation3-commit}
    return $\true$ // \label{htmsct-transformation3-return-true}\vspace{-1mm}
\end{lstlisting}
\end{framed}
    \vspace{-5mm}
	\caption{Transforming \sct$_O$: after completely eliminating accesses to fields of \op s created on the fast path.}
	\label{code-3path-htmsct-transformation3}
\end{figure*}
We now describe a transformation that completely eliminates all accesses to the $state$ fields of \op s created by transactions in $\sct_2$ (i.e., the last remaining accesses by transactions to fields of \op s).

We transform \sct$_2$ into a new procedure \sct$_3$, which appears in Figure~\ref{code-3path-htmsct-transformation3}.
First, the \cstep\ in $\sct_2$ is eliminated.
Whereas in $\sct_2$, we stored a pointer to the \op\ in $r.\info$ for each $r \in V$ at line~\ref{htmsct-transformation2-fcas}, we store a \textit{tagged pointer} to the \op\ at line~\ref{htmsct-transformation3-fcas} in $\sct_3$.
A tagged pointer is simply a pointer that has its least significant bit set to one.
Note that, on modern systems where pointers are word aligned, the least significant bit in a pointer to an \op\ will be zero.
Thus, the least significant bit in a tagged pointer allows processes to distinguish between a tagged pointer (which is stored in $r.\info$ by a transaction) from a regular pointer (which is stored in $r.\info$ by an invocation of $\sct_O$).
Line~\ref{htmsct-transformation3-abort} in $\sct_3$ is also updated to check for a tagged pointer in $r.\info$.

In order to deal with tagged pointers, we transform \llt$_O$ into new procedure called \llt$_{HTM}$, that is used instead of \llt$_O$ from here on.
Any time an invocation of \llt$_O$ would follow a pointer that was read from an $\info$ field $r.\info$, \llt$_{HTM}$ first checks whether the value $r\info$ read from the $\info$ field is a pointer or a tagged pointer.
If it is a pointer, then \llt$_{HTM}$ proceeds exactly as in \llt$_O$.
However, if $r\info$ is a tagged pointer, then \llt$_{HTM}$ proceeds as if it had seen an \op\ with \textit{state} \done\ (i.e., whose \sct\ has already returned \true).
We explain why this is correct.
If $r\info$ contains a tagged pointer, then it was written by a transaction $T$ that committed (since it changed shared memory) at line~\ref{htmsct-transformation3-commit} in \sct$_3$, just before returning \true.
Observe that, in \sct$_2$, the $state$ of the \op\ is set to \done\ just before \true\ is returned.
In other words, if not for this transformation, $T$ would have set the $state$ of its \op\ to \done.
So, clearly it is correct to treat $r\info$ as if it were an \op\ with $state = \done$.

Since this transformation simply changes the \textit{representation} of an \op\ $D$ with $state =$ \done\ that is created by a transaction (and does not change how the algorithm behaves when it encounters $D$), it preserves PROG.



\fakeparagraph{Eliminating the creation of \op s on the fast path}
Since transactions in $\sct_3$ are not helped, we would like to eliminate the \textit{creation} of \op s in transactions, altogether. 
However, since \op s are used as part of the \textit{freezing} mechanism in \sct$_O$ on the fallback path, we cannot simply eliminate the steps that freeze \rec s, or else transactions on the fast path will not synchronize with \sct$_O$ operations on the fallback path.
Consider an invocation $S$ of \sct$_O$ by a process $p$ that creates an \op\ $D$, and an invocation $L$ of \llt($r$) linked to $S$.
When $S$ uses CAS to freeze $r$ (by changing $r.\info$ from the value seen by $L$ to $D$), it interprets the success of the CAS to mean that $r$ has not changed since $L$ (relying on property P1).
If a transaction in \sct$_3$ changes $r$ without changing $r.\info$ (to a new value that has never before appeared in $r.\info$), then it would violate P1, rendering this interpretation invalid.
Thus, transactions in \sct$_3$($V, R, fld, new$) must change $r.\info$ to a new value, for each $r \in V$.


We transform \sct$_3$ into a new procedure \sct$_r$, which appears in Figure~\ref{code-3path-htmsct-transformation4}.
We now explain what a transaction $T$ in an invocation $S$ of \sct$_4$ by a process $p$ does instead of creating an \op\ and using it to freeze \rec s.
We give each process $p$ a \textit{tagged sequence number} \textit{tseq}$_p$, which consists of three bit fields: a tag-bit, a process name, and a sequence number. 
The tag-bit, which is the least significant bit, is always one.
This tag-bit distinguishes tagged sequence numbers from pointers to \op s (similar to tagged pointers, above).
The process name field of \textit{tseq}$_p$ contains $p$.
The sequence number is a non-negative integer that is initially zero.
Instead of creating a new \op\ (at line~\ref{htmsct-transformation3-new-op} in $\sct_3$), $S$ increments the sequence number field of \textit{tseq}$_p$.
Then, instead of storing a pointer to an \op\ in $r.\info$ for each $r \in V$ (at line~\ref{htmsct-transformation3-fcas} in \sct$_3$), $T$ stores $\textit{tseq}_p$.
(Line~\ref{htmsct-transformation3-abort} is also changed accordingly.)
The combination of the process name and sequence number bit fields ensure that whenever $T$ stores $\textit{tseq}_p$ in an $\info$ field, it is storing a value that has never previously been contained in that field.\footnote{Technically, with a finite word size it is possible for a sequence number to overflow and wrap around, potentially causing P1 to be violated. 
On modern systems with a 64-bit word size, we suggest representing a tagged sequence number using 1 tag-bit, 15 bits for the process name (allowing up to 32,768 concurrent processes) and 48 bits for the sequence number.
In order for a sequence number to experience wraparound, a \textit{single process} must then perform $2^{48}$ operations.
According to experimental measurements for several common data structures on high performance systems, this would take at least a decade of continuous updates.
Moreover, if wraparound is still a concern, one can replace the \fcas\ steps in \sct$_O$ with double-wide CAS instructions (available on most modern systems) which atomically operate on 128-bits.} 

\begin{figure*}[th]
\footnotesize
\def\pwidth{4cm}
\prepnewlisting
\vspace{-2mm}
\begin{framed}
\begin{lstlisting}[mathescape=true]
  //Private variable for process $p$: $\textit{tseq}_p$ \vspace{1mm}\hrule\vspace{1mm}
  //\sct$_4$($V, R, fld, new$) by process $p$
    //\dline{Let $\llresults$ be a pointer to a table in $p$'s private memory containing,}%
            {for each $r$ in $V$, the value of $r.\info$ read by $p$'s last \llt($r$)}
    //Let $old$ be the value for $fld$ returned by $p$'s last \llt($r$)\vspace{1.5mm}%

    //Begin hardware transaction
    //\textit{tseq}$_p$ := \textit{tseq}$_p$ + $2^{\lceil \log n \rceil}$ \hfill \com{increment $p$'s tagged sequence number}
    //\com \mbox{Freeze all \rec s in $V$ to protect their mutable fields from being changed by other \sct s}
    for each $r$ in $V \mbox{ enumerated in order}$ do
      //Let $r$\info\ be the pointer indexed by $r$ in $\llresults$
      if not $\cas$($r.\info,r\info,\textit{tseq}_p$) then //\sidecom{\textbf{\fcas}}\label{htmsct-transformation4-fcas}
        if $r.\info \neq \textit{tseq}_p$ then //Abort hardware transaction (explicitly)\vspace{1.5mm}%

    //\com Finished freezing \rec s
    //\com Finalize each $r \in R$, update $fld$, and unfreeze all $r \in $($V \setminus R$)
    for each $r \in R$ do $r.marked := \true$ //\sidecom{\textbf{\markstep}}
    //$\cas$($fld, old, new$) \sidecom{\textbf{\upcas}}\label{htmsct-transformation4-upcas}
    //Commit hardware transaction \label{htmsct-transformation4-commit2}
    return $\true$ //\vspace{-1mm}
\end{lstlisting}
\end{framed}
    \vspace{-5mm}
	\caption{Transforming \sct$_O$: after eliminating \op\ creation on the fast path.}
	\label{code-3path-htmsct-transformation4}
\end{figure*}


Observe that $\llt_{HTM}$ does not require any further modification to work with tagged sequence numbers, since it distinguishes between tagged sequence numbers and \op s using the tag-bit (the exact same way it distinguished between tagged pointers and pointers to \op s).
Moreover, it remains correct to treat tagged sequence numbers as if they are \op s with $state$ \done\ (for the same reason it was correct to treat tagged pointers that way).
Progress is preserved for the same reason as it was in the previous transformation: we are simply changing the \textit{representation} of \op s with $state =$ \done\ that are created by transactions.

Note that this transformation eliminates not only the \textit{creation} of \op s, but also the need to \textit{reclaim} those \op s.
Thus, it can lead to significant performance improvements.

\fakeparagraph{Simple optimizations}
\begin{figure*}[th]
\footnotesize
\def\pwidth{4cm}
\prepnewlisting
\vspace{-2mm}
\begin{framed}
\begin{lstlisting}[mathescape=true]
  //Private variable for process $p$: $\textit{tseq}_p$ \vspace{1mm}\hrule\vspace{1mm}
  //\sct$_5$($V, R, fld, new$) by process $p$
    //\dline{Let $\llresults$ be a pointer to a table in $p$'s private memory containing,}%
            {for each $r$ in $V$, the value of $r.\info$ read by $p$'s last \llt($r$)}
    //Let $old$ be the value for $fld$ returned by $p$'s last \llt($r$)\vspace{1.5mm}%

    //Begin hardware transaction
    //\textit{tseq}$_p$ := \textit{tseq}$_p$ + $2^{\lceil \log n \rceil}$ \hfill \com{increment $p$'s tagged sequence number}
    //\com \mbox{Freeze all \rec s in $V$ to protect their mutable fields from being changed by other \sct s}
    for each $r$ in $V \mbox{ enumerated in order}$ do
      //Let $r$\info\ be the pointer indexed by $r$ in $\llresults$
      if $r.\info = r\info$ then $r.\info := \textit{tseq}_p$ //\label{htmsct-transformation5-freezing}
      else //Abort hardware transaction (explicitly) \vspace{1.5mm}%

    //\com Finished freezing \rec s
    //\com Finalize each $r \in R$, update $fld$, and unfreeze all $r \in $($V \setminus R$)
    for each $r \in R$ do $r.marked := \true$ //\sidecom{\textbf{\markstep}}
    if $fld = old$ then $fld := new$ //\label{htmsct-transformation5-update}
    //Commit hardware transaction
    return $\true$ //\vspace{-1mm}
\end{lstlisting}
\end{framed}
    \vspace{-4.5mm}
	\caption{Transforming \sct$_O$: after replacing CAS with sequential code and optimizing.}
	\label{code-3path-htmsct-transformation5}
\end{figure*}
Since any code executed inside a transaction is atomic, we are free to replace atomic synchronization primitives inside a transaction with sequential code, and reorder the transaction's steps in any way that does not change its sequential behaviour. 
We now describe how to transform \sct$_4$ by performing two simple optimizations.

For the first optimization, we replace each invocation of CAS($x,$ $o,$ $n$) with sequential code: \textbf{if} $x = 0$ \textbf{then} $x := n, result := \true$ \textbf{else} $result := \false$.
If the CAS is part of a condition for an if-statement, then we execute this code just before the if-statement, and replace the invocation of CAS with $result$.
We then eliminate any \textit{dead code} that cannot be executed.
Figure~\ref{code-3path-htmsct-transformation5} shows the transformed procedure, \sct$_5$.

More concretely, in place of the CAS at line~\ref{htmsct-transformation4-fcas} in \sct$_4$, we do the following.
First, we check whether $r.\info = r\info$.
If so, we set $r.\info := \textit{tseq}_p$ and continue to the next iteration of the loop.
Suppose not.
If we were naively transforming the code, then the next step would be to check whether $r.\info$ contains $\textit{tseq}_p$.
However, 
$p$ is the only process that can write $\textit{tseq}_p$, and it only writes $\textit{tseq}_p$ just before continuing to the next iteration.
Thus, $r.\info$ cannot possibly contain $\textit{tseq}_p$ in this case, which makes it unnecessary to check whether $r.\info = \textit{tseq}_p$.
Therefore, we execute the \textit{else}-case, and explicitly abort the transaction.
Observe that, if \sct$_5$ is used to replace \sct$_1$ in Figure~\ref{code-3path-htmsct-usage}, then this explicit abort will cause \sct\ to return \false\ (right after it jumps to the onAbort label).
In place of the CAS at line~\ref{htmsct-transformation4-upcas} in \sct$_4$, we can simply check whether $fld$ contains $old$ and, if so, write $new$ into $fld$.

In fact, it is not necessary to check whether $fld$ contains $old$, because the transaction will have aborted if $fld$ was changed after $old$ was read from it.
We explain why.
Let $S$ be an invocation of \sct$_5$ (in Figure~\ref{code-3path-htmsct-transformation5}) by a process $p$, and let $r$ be the \rec\ that contains $fld$.
Suppose $S$ executes line~\ref{htmsct-transformation5-update} in \sct$_5$, where it checks whether $fld = old$.
Before invoking $S$, $p$ performs an invocation $L$ of \llt($r$) linked to $S$.
Subsequently, $p$ reads $old$ while performing $S$.
After that, $p$ freezes $r$ while performing $S$.
If $r$ changes after $L$, and before $p$ executes line~\ref{htmsct-transformation5-freezing}, then $p$ will see $r.\info \neq r\info$ when it executes line~\ref{htmsct-transformation5-freezing} (by property P1, which has been preserved by our transformations).
Consequently, $p$ will fail to freeze $r$, and $S$ will perform an explicit abort and return \false, so it will \textit{not} reach line~\ref{htmsct-transformation5-update}, which contradicts our assumption (so this case is impossible).
On the other hand, if $r$ changes after $p$ executes line~\ref{htmsct-transformation5-freezing}, and before $p$ executes line~\ref{htmsct-transformation5-update}, then the transaction will abort due to a data conflict (detected by the HTM system).
Therefore, when $p$ executes line~\ref{htmsct-transformation5-update}, $fld$ must contain $old$.

\begin{figure*}[t]
\footnotesize
\def\pwidth{4cm}
\prepnewlisting
\begin{framed}
\begin{lstlisting}[mathescape=true]
  //Private variable for process $p$: $\textit{tseq}_p$ \vspace{1mm}\hrule\vspace{1mm}
  //\sct$_{HTM}$($V, R, fld, new$) by process $p$
    //\dline{Let $\llresults$ be a pointer to a table in $p$'s private memory containing,}%
            {for each $r$ in $V$, the value of $r.\info$ read by $p$'s last \llt($r$)}
    //Let $old$ be the value for $fld$ returned by $p$'s last \llt($r$)\vspace{1.5mm}%

    //Begin hardware transaction
    //\textit{tseq}$_p$ := \textit{tseq}$_p$ + $2^{\lceil \log n \rceil}$ \hfill \com{increment $p$'s tagged sequence number} \label{htmsct-tseq-increment}
    for each $r \in V$ do //\hfill\com abort if any $r \in V$ has changed since the linked \llt($r$) \label{htmsct-abort}
      //Let $r$\info\ be the pointer indexed by $r$ in $\llresults$ \label{htmsct-rinfo}
      if $r.\info \neq r\info$ then //Abort hardware transaction (explicitly)
    for each $r \in V$ do $r.\info := \textit{tseq}_p$ //\hfill\com change $r.\info$ to a new value, for each $r \in V$ \label{htmsct-change-info} \label{htmsct-fcas}
    for each $r \in R$ do $r.marked := \true$//\hfill\com mark each $r \in R$ (so it will be finalized) \label{htmsct-mark}
    //write $new$ to the field pointed to by $fld$ \hfill\com perform the update \label{htmsct-write-fld}
    //Commit hardware transaction
    return $\true$ //\vspace{-1mm}
\end{lstlisting}
\end{framed}
    \vspace{-5mm}
	\caption{Final implementation of \sct$_{HTM}$.}
	\label{code-3path-htmsct}
\end{figure*}

For the second optimization, we split the loop in Figure~\ref{code-3path-htmsct-transformation5} into two.
The first loop contains all of the steps that check whether $r.\info = r\info$, and the second loop contains all of the steps that set $r.\info := \textit{tseq}_p$.
This way, all of the writes to $r.\info$ occur after all of the reads and if-statements.
The advantage of delaying writes for as long as possible in a transaction is that it reduces the probability of the transaction 
causing other transactions to abort. 
As a minor point, whereas the loop in \sct$_O$ iterated over the elements of the sequence $V$ in a particular order to guarantee progress, it is not necessary to do so here, since progress is guaranteed by the fallback path, not the fast path.
This final transformation yields the code in Figure~\ref{code-3path-newscx}.
Clearly, it does not affect correctness or progress.

\section{Accelerated template implementations} \label{sec-3path-algs}

\fakeparagraph{The \textit{2-path con} algorithm}
We now use our HTM-based \llt\ and \sct\ to obtain an HTM-based implementation of a template operation $O$.
The fallback path for $O$ is simply a lock-free implementation of $O$ using \llt$_O$ and \sct$_O$.
The fast path for $O$ starts a transaction, then performs the same code as the fallback path, except that it uses the HTM-based \llt\ and \sct. 
Since the \textit{entire operation} is performed inside a transaction, we can optimize the invocations of \sct$_{HTM}$ that are performed by $O$ as follows.
Lines~\ref{newscx-scxhtm-xbegin} and~\ref{newscx-scxhtm-commit} can be eliminated, since \sct$_{HTM}$ is already running inside a large transaction. 
Additionally, lines~\ref{newscx-scxhtm-freezing-loop-start}-\ref{newscx-scxhtm-freezing-loop-end} can be eliminated, since the transaction will abort due to a data conflict if $r.\info$ changes after it is read in the (preceding) linked invocation of \llt($r$), and before the transaction commits.
The proof of correctness and progress for \textit{2-path con} follows immediately from the proof of the original template and the proof of the HTM-based \llt\ and \sct\ implementation.

Note that it is not necessary to perform the entire operation in a single transaction.
In Section~\ref{sec-3path-search-outside}, we describe a modification that allows a read-only \textit{searching} prefix of the operation to be performed before the transaction begins. 

\fakeparagraph{The \textit{TLE} algorithm}
%
To obtain a TLE implementation of an operation $O$, we simply take \textit{sequential code} for $O$ and wrap it in a transaction on the fast path.
The fallback path acquires and releases a global lock instead of starting and committing a transaction, but otherwise executes the same code as the fast path.
To prevent the fast path and fallback path from running concurrently, transactions on the fast path start by reading the lock state and aborting if it is held.
An operation attempts to run on the fast path up to \textit{AttemptLimit} times (waiting for the lock to be free before each attempt) before resorting to the fallback path.
The correctness of TLE is trivial.
Note, however, that TLE only satisfies deadlock-freedom (not lock-freedom).

%

\fakeparagraph{The \textit{2-path $\overline{con}$} algorithm}
\later{talk about a counter object instead of fetch-and-increment, because the latter is stronger than we need. the former can be implemented using only registers.}
We can improve concurrency on the fallback path and guarantee lock-freedom by using a lock-free algorithm on the fallback path, and a global fetch-and-increment object $F$ instead of a global lock.
Consider an operation $O$ implemented with the tree update template.
We describe a \textit{2-path $\overline{con}$} implementation of $O$.
The fallback path increments $F$, then executes the lock-free tree update template implementation of $O$, and finally decrements $F$.
The fast path executes \textit{sequential code} for $O$ in a transaction.
To prevent the fast path and fallback path from running concurrently, transactions on the fast path start by reading $F$ and aborting if it is nonzero.
An operation attempts to run on the fast path up to \textit{AttemptLimit} times (waiting for $F$ to become zero before each attempt) before resorting to the fallback path.

Recall that operations implemented using the tree update template can only change a single pointer atomically (and can perform multiple changes atomically only by creating a connected set of new nodes that reflect the desired changes).
Thus, each operation on the fallback path simply creates new nodes and changes a single pointer (and assumes that all other operations also behave this way).
However, since the fast path and fallback path do not run concurrently, the fallback path does \textit{not} impose this requirement on the fast path.
Consequently, the fast path can make (multiple) direct changes to nodes. 
Unfortunately, as we described above, this algorithm can still suffer from concurrency bottlenecks.


\fakeparagraph{The \textit{3-path} algorithm}
One can think of the \textit{3-path} algorithm as a kind of hybrid between the \textit{2-path con} and \textit{2-path $\overline{con}$} algorithms that obtains their benefits while avoiding their downsides.
Consider an operation $O$ implemented with the tree update template.
We describe a 3-path implementation of $O$.
As in \textit{2-path $\overline{con}$}, there is a global fetch-and-increment object $F$, and the fast path executes \textit{seqeuential code} for $O$ in a transaction.
The middle path and fallback path behave like the fast path and fallback path in the \textit{2-path con} algorithm, respectively. 
Each time an operation begins (resp., stops) executing on the fallback path, it increments (resp., decrements) $F$.
(If the scalability of fetch-and-increment is of concern, then a \textit{scalable non-zero indicator} object~\cite{ellen2007snzi} can be used, instead. Alternatively, one could use a \textit{counter} object, which is weaker and can be implemented using only registers.)
This prevents the fast and fallback paths from running concurrently.
As we described above, operations begin on the fast path, and move to the middle path after \textit{FastLimit} attempts, or if they see $F \neq 0$.
Operations move from the middle path to the fallback path after \textit{MiddleLimit} attempts.
Note that an operation never waits for the fallback path to become empty---it simply moves to the middle path.

Since the fast path and fallback path do not run concurrently, the fallback path does not impose any overhead on the fast path, except checking if $F = 0$ (offering low overhead for light workloads).
Additionally, when there are operations running on the fallback path, hardware transactions can continue to run on the middle path (offering high concurrency for heavy workloads).

\fakeparagraph{Correctness and progress for \textit{3-path}}
The correctness argument is straightforward.
The goal is to prove that all template operations are linearizable, regardless of which path they execute on.
Recall that the fallback path and middle path behave like the fast path and fallback path in \textit{2-path con}. 
It follows that, if there are no operations on the fast path, then the correctness of operations on the middle path and fallback path is immediate from the correctness of \textit{2-path con}. 
Of course, whenever there is an operation executing on the fallback path, no operation can run on the fast path.
Since operations on the fast path and middle path run in transactions, they are atomic, and any conflicts between the fast path and middle path are handled automatically by the HTM system.
Therefore, template operations are linearizable.

The progress argument for \textit{3-path} relies on three assumptions.
\begin{compactenum}[\bf {A}1.]
    \item The sequential code for an operation executed on the fast path must terminate after a finite number of steps if it is run on a static tree (which does not change during the operation).
    \item In an operation executed on the middle path or fallback path, the search phase must terminate after a finite number of steps if it is run on a static tree.
    \item In an operation executed on the middle path or fallback path, the update phase can modify only a finite number of nodes.
\end{compactenum}

\smallskip

We give a simple proof that \textit{3-path} satisfies lock-freedom. 
To obtain a contradiction, suppose there is an execution in which after some time $t$, some process takes infinitely many steps, but no operation terminates.
Thus, the tree does not change after $t$.
We first argue that no process takes infinitely many steps in a transaction $T$.
If $T$ occurs on the fast path, then A1 guarantees it will terminate.
If $T$ occurs on the middle path, then A2 and A3 guarantee that it will terminate.
Therefore, eventually, processes only take steps on the fallback path.
Progress then follows from the fact that the original tree update template implementation (our fallback path) is lock-free.

\section{Example data structures} \label{sec-3path-ds}

We used two data structures to study the performance of our accelerated template implementations: an unbalanced BST, and a relaxed ($a,b$)-tree.
The BST is similar to the chromatic tree in~\cite{Brown:2014}, but with no rebalancing.
The relaxed ($a,b$)-tree is a concurrency-friendly generalization of a B-tree that is based on the work of Jacobson and Larsen~\cite{JL01abtrees}.
In this section, we give a more detailed description of these data structures, and give additional details on their \textit{3-path} implementations. 

Each data structure implements the ordered dictionary ADT, which stores a set of keys, and associates each key with a value.
An ordered dictionary offers four operations: \textsc{Insert}($key, value$), \textsc{Delete}($key$), \textsc{Search}($key$) and \textsc{RangeQuery}($lo, hi$).

Both data structures are \textit{leaf-oriented} (also called \textit{external}), which means that all of the keys in the dictionary are stored in the leaves of the tree, and internal nodes contain \textit{routing} keys which simply direct searches to the appropriate leaf.
This is in contrast to \textit{node-oriented} or \textit{internal} trees, in which internal nodes also contain keys in the set.

\subsection{Unbalanced BST}


\fakeparagraph{Fallback path}
\begin{figure*}
    \centering
    \includegraphics[width=\linewidth]{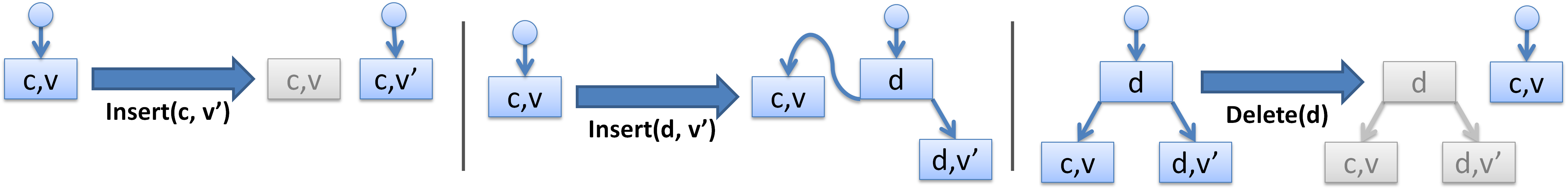}
    \caption{Fallback path operations for the unbalanced BST.}
    \label{fig-lfbst-ops}
\end{figure*}
The fallback path consists of a lock-free implementation of the operations in Figure~\ref{fig-lfbst-ops} using the (original) tree update template. 
As required by the template, these operations change child pointers, but do not change the key or value fields of nodes directly.
Instead, to replace a node's key or value, the node is replaced by a new copy.
If $key$ is not already in the tree, then \textit{Insert}($key, value$) inserts a new leaf and internal node.
Otherwise, \textit{Insert}($key, value$) replaces the leaf containing $key$ with a new leaf that contains the updated value.
\textit{Delete}($key$) replaces the leaf $l$ being deleted and its parent with a new copy of the sibling of $l$.

%
It may seem strange that \textit{Delete} creates a new copy of the deleted leaf's sibling, instead of simply reusing the existing sibling (which is not changed by the deletion).
This comes from a requirement of the tree update template: each invocation of \sct($V, R, fld, new$) must change the field $fld$ to a value that it has \textit{never previously contained}.
%
This requirement is motivated by a particularly tricky aspect of lock-free programming: avoiding the \textit{ABA problem}.
The ABA problem occurs when a process $p$ reads a memory location $x$ and sees value $A$, then performs a CAS on $x$ to change it from $A$ to $C$, and \textit{interprets} the success of this CAS to mean that $x$ has not changed between when $p$ read $x$ and performed the CAS on it.
In reality, after $p$ read $x$ and before it performed the CAS, another process $q$ may have changed $x$ to $B$, and then back to $A$, rendering $p$'s interpretation 
invalid.
In practice, the ABA problem can result in data structure operations being applied multiple times, or lost altogether.
The ABA problem cannot occur if each successful CAS on a field stores a value that has never previously been contained in the field (since, then, $q$ cannot change $x$ from $B$ back to $A$).
So, in the template, the ABA problem is avoided by having each operation use \sct\ to store \textit{a pointer to a newly created node} (which cannot have previously been contained in any field).

\fakeparagraph{Middle path} 
The middle path is the same as the fallback path, except that each operation is performed in a large transaction, and the HTM-based implementation of \llt\ and \sct\ is used instead of the original implementation.

\fakeparagraph{Fast path}
\begin{figure*}
    \centering
    \includegraphics[width=\linewidth]{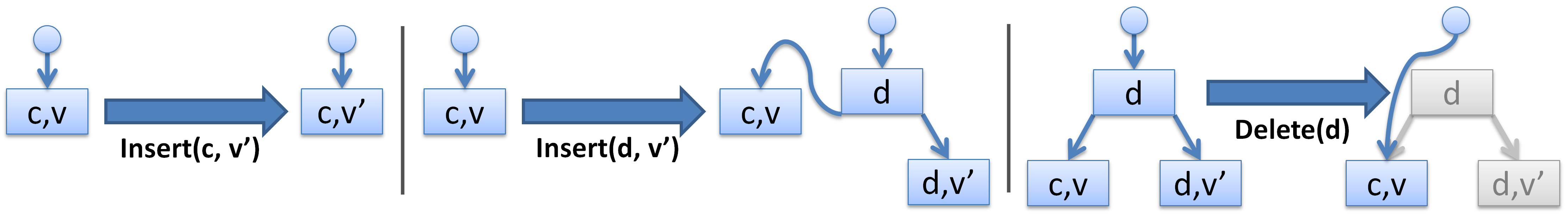}
    \caption{
        Fast path operations for the unbalanced BST.
        (\textit{Insert}($d, v'$) is the same as on the fallback path.)
    }
    \label{fig-lfbst-ops-fastpath}
\end{figure*}
The fast path is a sequential implementation of the BST, where each operation is executed in a transaction.
Figure~\ref{fig-lfbst-ops-fastpath} shows the insertion and deletion operations on the fast path.
Unlike on the fallback path, operations on the fast path directly modify the keys and values of nodes, and, hence, can avoid creating nodes in some situations.
If $key$ is already in the tree, then \textit{Insert}($key, value$) directly changes the value of the leaf that contains $key$.
Otherwise, \textit{Insert}($key, value$) creates a new leaf and internal node and attaches them to the tree.
\textit{Delete}($key$) changes a pointer to remove the leaf containing $key$ and its parent from the tree.

\fakeparagraph{How the fast path improves performance}
The first major performance improvement on the fast path comes from a reduction in node creation.
Each invocation of \textit{Insert}($key, value'$) that sees $key$ in the tree can avoid creating a new node by writing $value'$ directly into the node that already contains $key$.
In contrast, a new node had to be created on the middle path, since the middle path runs concurrently with the fallback path, which assumes that the keys and values of nodes do not change.
Additionally, each invocation of \textit{Delete} that sees $key$ in the tree can avoid creating a new copy of the sibling of the deleted leaf.
This optimization was not possible on the middle path, because the fallback path assumes that each successful operation writes a pointer to a newly created node.
The second major improvement comes from the fact that reads and writes suffice where invocations of LLX and SCX were needed on the other paths.

\subsection{Relaxed (a,b)-tree}
%
%
The relaxed ($a,b$)-tree~\cite{JL01abtrees} is a generalization of a B-tree.
Larsen introduced the relaxed ($a,b$)-tree as a sequential data structure that was well suited to fine-grained locking.
Internal nodes contain up to $b-1$ \textit{routing} keys, and have one more child pointer than the number of keys.
Leaves contain up to $b$ key-value pairs (which are in the dictionary).
(Values may be pointers to large data objects.)
The \textit{degree} of an internal node (resp. leaf) is the number of pointers (resp. keys) it contains.
When there are no ongoing updates (insertions and deletions) in a relaxed ($a,b$)-tree, all leaves have the same depth, and nodes have \textit{degree} at least $a$ and at most $b$, where $b \ge 2a-1$.
Maintaining this balance condition requires rebalancing steps similar to the \textit{splits} and \textit{joins} of B-trees.
(See~\cite{JL01abtrees} for further details on the rebalancing steps.)

\fakeparagraph{Fallback path}
The fallback path consists of a lock-free implementation of the relaxed ($a,b$)-tree operations using the (original) tree update template.
If $key$ is in the tree, then \textit{Insert}($key, value$) replaces the leaf containing $key$ with a new copy that contains ($key, value$).
Suppose $key$ is not in the tree.
Then, \textit{Insert} finds the leaf $u$ where the key should be inserted.
If $u$ is not full (has degree less than $b$), then it is replaced with a new copy that contains ($key, value$).
Otherwise, $u$ is replaced by a subtree of three new nodes: one parent and two children.
The two new children evenly share the key-value pairs of $u$ and ($key, value$).
The new parent $p$ contains only a single routing key and two pointers (to the two new children), and is \textit{tagged}, which indicates that the subtree rooted at $p$ is too tall, and rebalancing should be performed to shrink its height.
\textit{Delete}($key$) replaces the leaf containing $key$ with a new copy $new$ that has $key$ deleted.
If the degree of $new$ is smaller than $a$, then rebalancing must be performed.

\fakeparagraph{Middle path}
This path is obtained from the fallback path the same way as in the unbalanced BST.

\fakeparagraph{Fast path}
The fast path is a sequential implementation of a relaxed ($a,b$)-tree whose operations are executed inside transactions.
Like the external BST, the major performance improvement over the middle path comes from the facts that (1) operations create fewer nodes, and (2) reads and writes suffice where LLX and SCX were needed on the other paths.
In particular, \textit{Insert}($key, value$) and \textit{Delete}($key$) simply directly modify the keys and values of leaves, instead of creating new nodes, except in the case of an \textit{Insert} into a full node $u$.
In that case, two new nodes are created: a parent and a sibling for $u$.
(Recall that this case resulted in the creation of three new nodes on the fallback path and middle path.)
Note that reducing node creation is more impactful for the relaxed ($a,b$)-tree than for the unbalanced BST, since nodes are much larger.

As a minor point, we found that it was faster in practice to perform rebalancing steps by creating new nodes, and simply replacing the old nodes with the new nodes that reflect the desired change (instead of rebalancing by directly changing the keys, values and pointers of nodes).

\section{Experiments} \label{sec-3path-exp}
We used two different Intel systems for our experiments: 
a dual-socket 12-core E7-4830 v3 with hyperthreading for a total of 48 hardware threads (running Ubuntu 14.04LTS), and a dual-socket 18-core E5-2699 v3 with hyperthreading for a total of 72 hardware threads (running Ubuntu 15.04).
Each machine had 128GB of RAM.
We used the scalable thread-caching allocator (tcmalloc) from the Google perftools library.
All code was compiled on GCC 4.8+ with arguments \texttt{-std=c++0x -O2 -mcx16}.
(Using the higher optimization level \texttt{-O3} did not significantly improve performance for any algorithm, and decreased performance for some algorithms.)
On both machines, we \textit{pinned} threads such that we saturate one socket before scheduling any threads on the other.

\fakeparagraph{Data structure parameters.}
Recall that nodes in the relaxed $(a,b)$-tree contain up to $b$ keys, and, when there are no ongoing updates, they contain at least $a$ keys (where $b \ge 2a-1$).
In our experiments, we fix $a=6$ and $b=16$.
With $b = 16$, each node occupies four consecutive cache lines.
Since $b \ge 2a-1$, with $b=16$, we must have $a \le 8$.
We chose to make $a$ slightly smaller than 8 in order to exploit a performance tradeoff: a smaller minimum degree may slightly increase depth, but decreases the number of rebalancing steps that are needed to maintain balance.

\fakeparagraph{Template implementations studied}
We implemented each of the data structures with four different template implementations: \textit{3-path}, \textit{2-path con}, 
\textit{TLE} and the original template implementation, which we call \textit{Non-HTM}.
(\textit{2-path $\overline{con}$} is omitted, since it performed similarly to TLE, and cluttered the graphs.)
%
The \textit{2-path con} and \textit{TLE} implementations perform up to 20 attempts on the fast path before resorting to the fallback path.
\textit{3-path} performs up to 10 attempts (each) on the fast path and middle path.
We implemented memory reclamation using DEBRA~\cite{Brown:2015}, an epoch based reclamation scheme.
A more efficient way to reclaim memory for \textit{3-path} is proposed in Section~\ref{sec-3path-memrecl}


\subsection{Light vs. Heavy workloads}

\fakeparagraph{Methodology}
We study two workloads: in \textbf{light}, $n$ processes perform updates (50\% insertion and 50\% deletion), and in \textbf{heavy}, $n-1$ processes perform updates, and one thread performs 100\% range queries (RQs).
For each workload and data structure implementation, and a variety of thread counts, we perform a set of five randomized trials.
In each trial, $n$ processes perform either updates or RQs (as appropriate for the workload) for one second, and counted the number of completed operations.
Updates are performed on keys
\begin{figure}[tb]
\centering
    \setlength\tabcolsep{0pt}
    \begin{tabular}{m{0.02\textwidth}m{0.22\textwidth}m{0.22\textwidth}}
        &
        \multicolumn{2}{c}{
            \fcolorbox{black!80}{black!40}{\parbox{\dimexpr 0.44\textwidth-2\fboxsep-2\fboxrule}{\centering\textbf{2x 24-thread Intel E7-4830 v3}}}
        }
        \\
        &
        \fcolorbox{black!50}{black!20}{\parbox{\dimexpr \linewidth-2\fboxsep-2\fboxrule}{\centering {\small Unbalanced BST (LLX/SCX)\\ Updates w/key range [$0,10^4$)}}} &
        \fcolorbox{black!50}{black!20}{\parbox{\dimexpr \linewidth-2\fboxsep-2\fboxrule}{\centering {\small (a, b)-tree (LLX/SCX)\\ Updates w/key range [$0,10^6$)}}}
        \\
        \vspace{-3mm}\rotatebox{90}{{\small\textbf{Light} workload}} &
        \includegraphics[width=\linewidth]{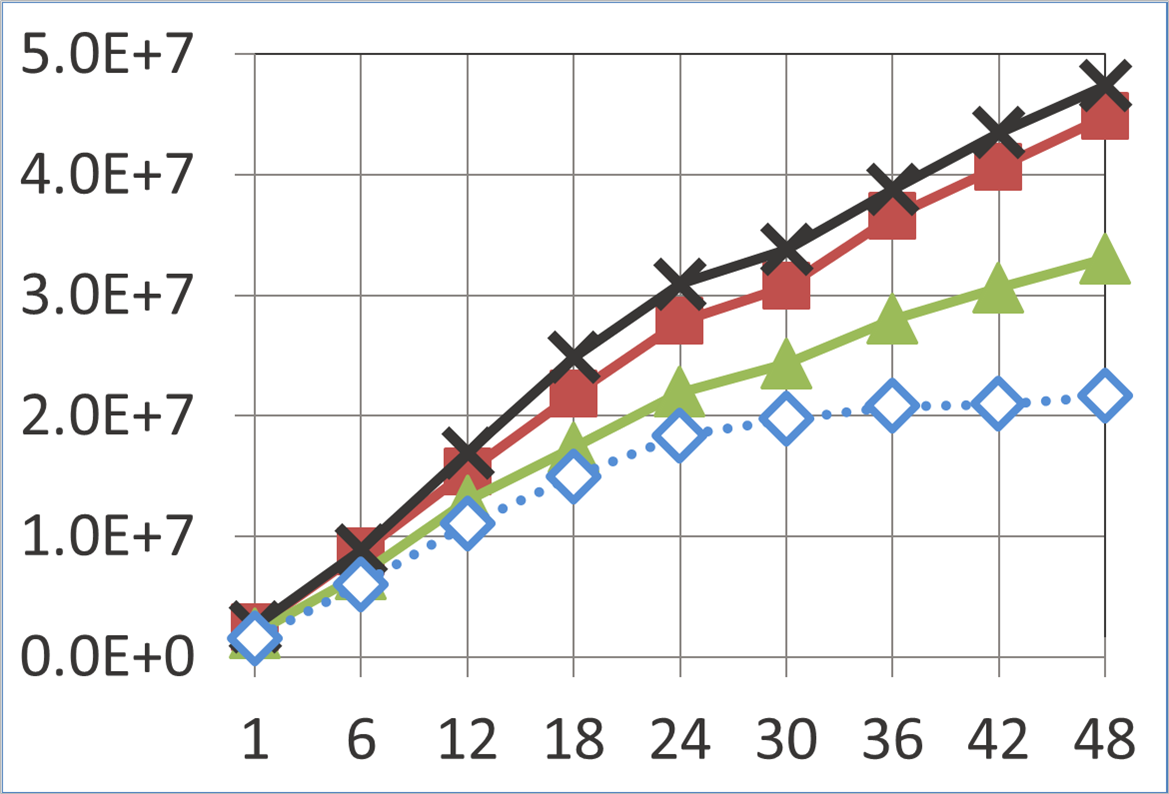} &
        \includegraphics[width=\linewidth]{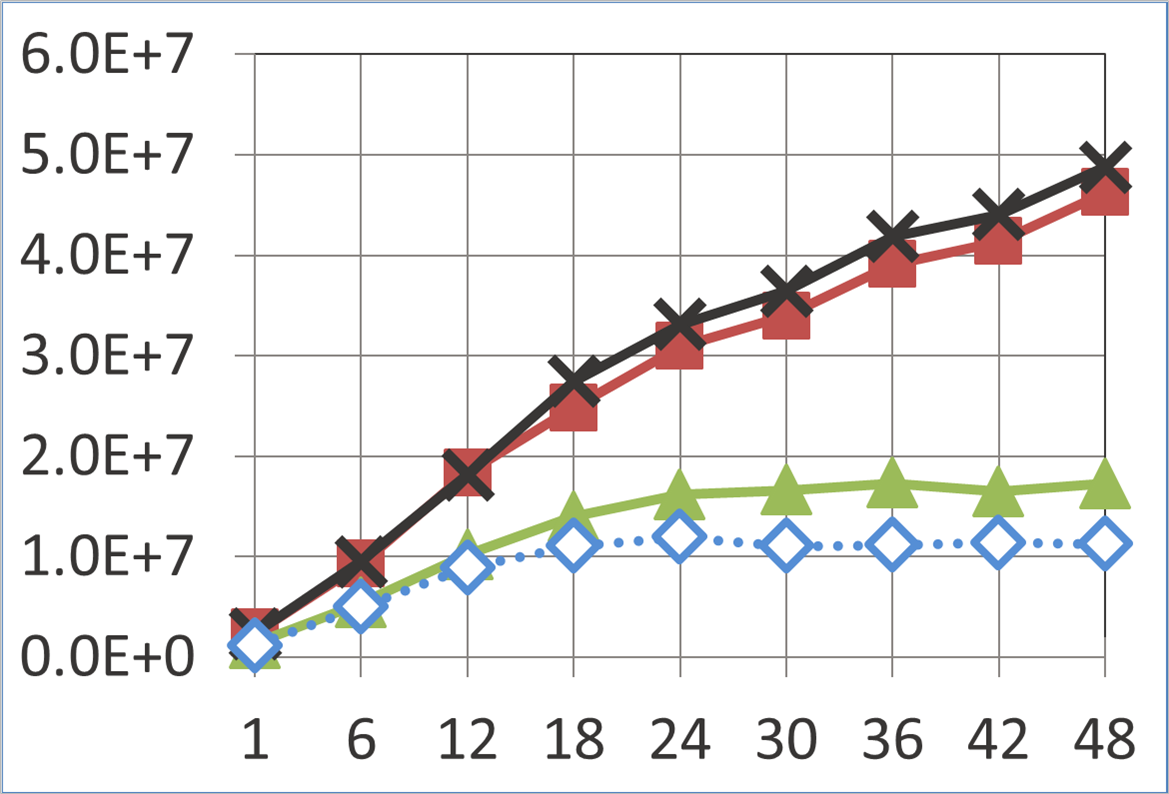}
        \\
        \vspace{-4.5mm}\rotatebox{90}{{\small\textbf{Heavy} workload}} &
        \vspace{-4.5mm}\includegraphics[width=\linewidth]{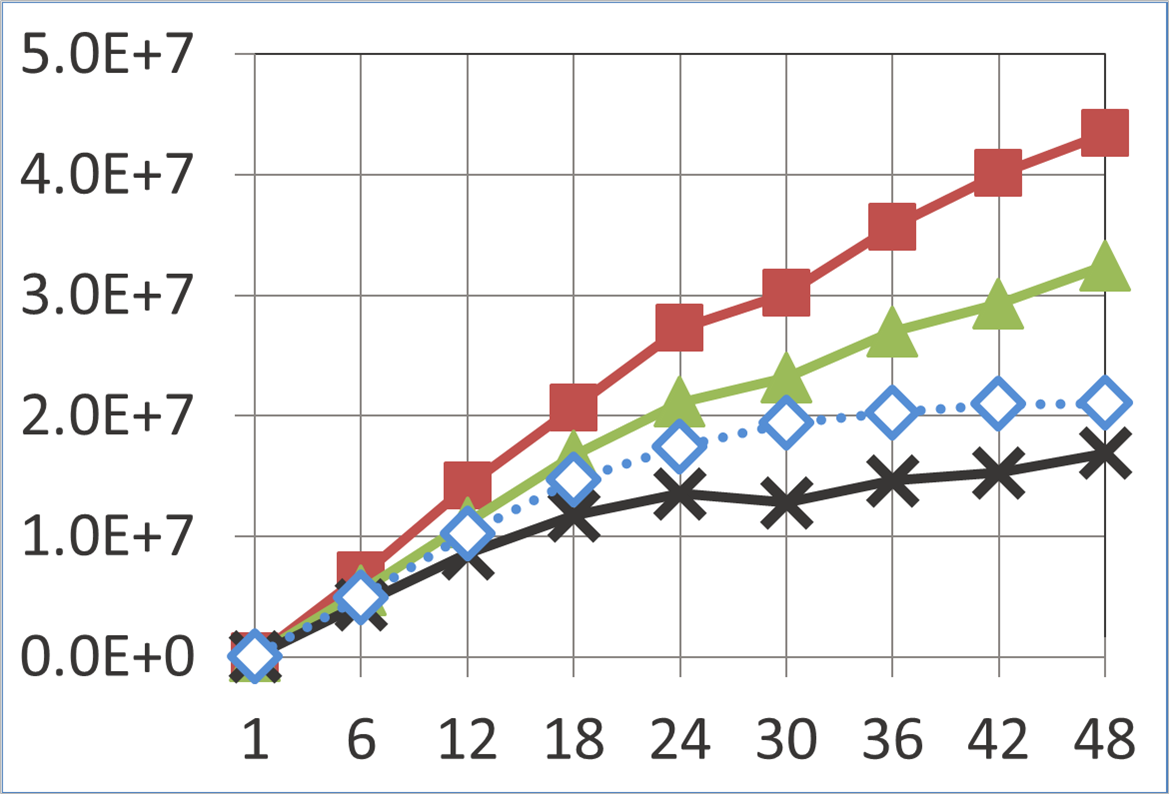} &
        \vspace{-4.5mm}\includegraphics[width=\linewidth]{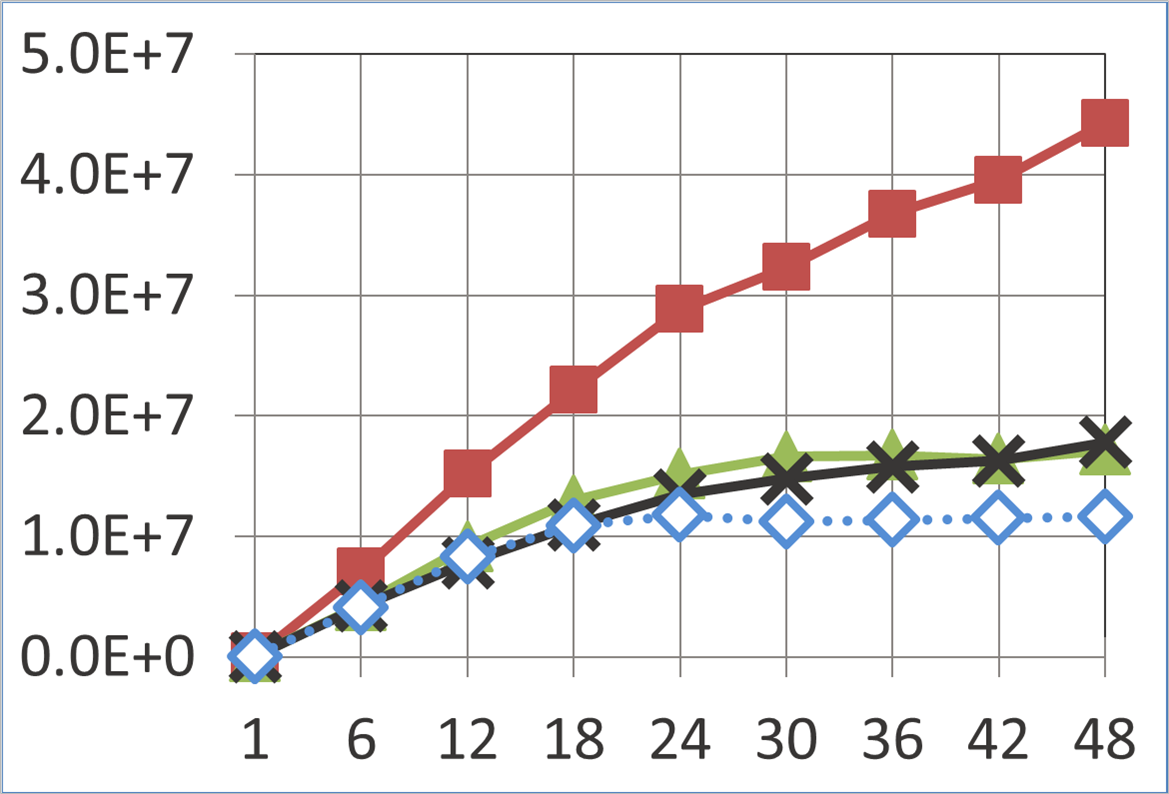}
        \\
    \end{tabular}

    \vspace{-1mm}
    \hspace{0.02\textwidth}\includegraphics[width=0.8\linewidth]{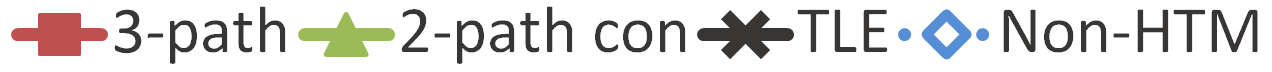}
    \caption{Results (48-thread system) showing throughput (operations per second) vs. concurrent processes.}
    \label{fig-graphs}
\end{figure}
drawn uniformly randomly from a fixed key range [$0, K$).
RQs are performed on ranges [$lo, lo+s$) where $lo$ is uniformly random in [$0, K$) and $s$ is chosen, according to a probability distribution described below, from [$1, 1000$] for the BST and [$1, 10000$] for the ($a,b$)-tree.
(We found that nodes in the ($a,b$)-tree contained approximately 10 keys, on average, so the respective maximum values of $s$ for the BST and ($a,b$)-tree resulted in range queries returning keys from approximately the same number of nodes in both data structures.)
To ensure that we are measuring steady-state performance, at the start of each trial, the data structure is prefilled by having threads perform 50\% insertions and 50\% deletions on uniform keys until the data structure contains approximately half of the keys in [$0, K$).

We verified the correctness of each data structure after each trial by 
computing \textit{key-sum hashes}.
Each thread maintains the sum of all keys it successfully inserts, minus the sum of all keys it successfully deletes.
At the end of the trial, the total of these sums over all threads must match the sum of keys in the tree. 

\fakeparagraph{Probability distribution of $s$}
We chose the probability distribution of $s$ to produce many small RQs, and a smaller number of very large ones.
To achieve this, we chose $s$ to be $\lfloor x^2 S \rfloor + 1$, where $x$ is a uniform real number in [$0, 1$), and $S = 1000$ for the BST and $S = 10000$ for the ($a,b$)-tree.
By squaring $x$, we bias the uniform distribution towards zero, creating a larger number of small RQs. 

\fakeparagraph{Results}
We briefly discuss the results from the 48 thread machine, which appear in Figure~\ref{fig-graphs}.
The BST and the relaxed ($a,b$)-tree behave fairly similarly.
Since the ($a,b$)-tree has large nodes, it benefits much more from a low-overhead fast path (in \textit{TLE} or \textit{3-path}) which can avoid creating new nodes during updates.
In the light workloads, \textit{3-path} performs significantly better than \textit{2-path con} (which has more overhead) and approximately as well as \textit{TLE}.
On average, the \textit{3-path} algorithms completed 2.1x as many operations as their \textit{non-HTM} counterparts (and with 48 concurrent processes, this increases to 3.0x, on average).
In the heavy workloads, \textit{3-path} significantly outperforms \textit{TLE} (completing 2.0x as many operations, on average), which suffers from \textit{excessive waiting}. 
Interestingly, \textit{3-path} is also significantly faster than \textit{2-path con} in the heavy workloads.
This is because, even though RQs are always being performed, some RQs can succeed on the fast path, so many update operations can still run on the fast path in \textit{3-path}, where they incur much less overhead (than they would in \textit{2-path con}).

\begin{figure}[tb]
    \centering
    \setlength\tabcolsep{0pt}
    \begin{tabular}{m{0.02\textwidth}m{0.22\textwidth}m{0.22\textwidth}}
        &
        \multicolumn{2}{c}{
            \fcolorbox{black!80}{black!40}{\parbox{\dimexpr 0.44\textwidth-2\fboxsep-2\fboxrule}{\centering\textbf{2x 36-thread Intel E5-2699 v3}}}
        }
        \\
        &
        \fcolorbox{black!50}{black!20}{\parbox{\dimexpr \linewidth-2\fboxsep-2\fboxrule}{\centering {\small Unbalanced BST (LLX/SCX)\\ Updates w/key range [$0,10^4$)}}} &
        \fcolorbox{black!50}{black!20}{\parbox{\dimexpr \linewidth-2\fboxsep-2\fboxrule}{\centering {\small (a, b)-tree (LLX/SCX)\\ Updates w/key range [$0,10^6$)}}}
        \\
        \vspace{-3mm}\rotatebox{90}{{\small\textbf{Light} workload}} &
        \includegraphics[width=\linewidth]{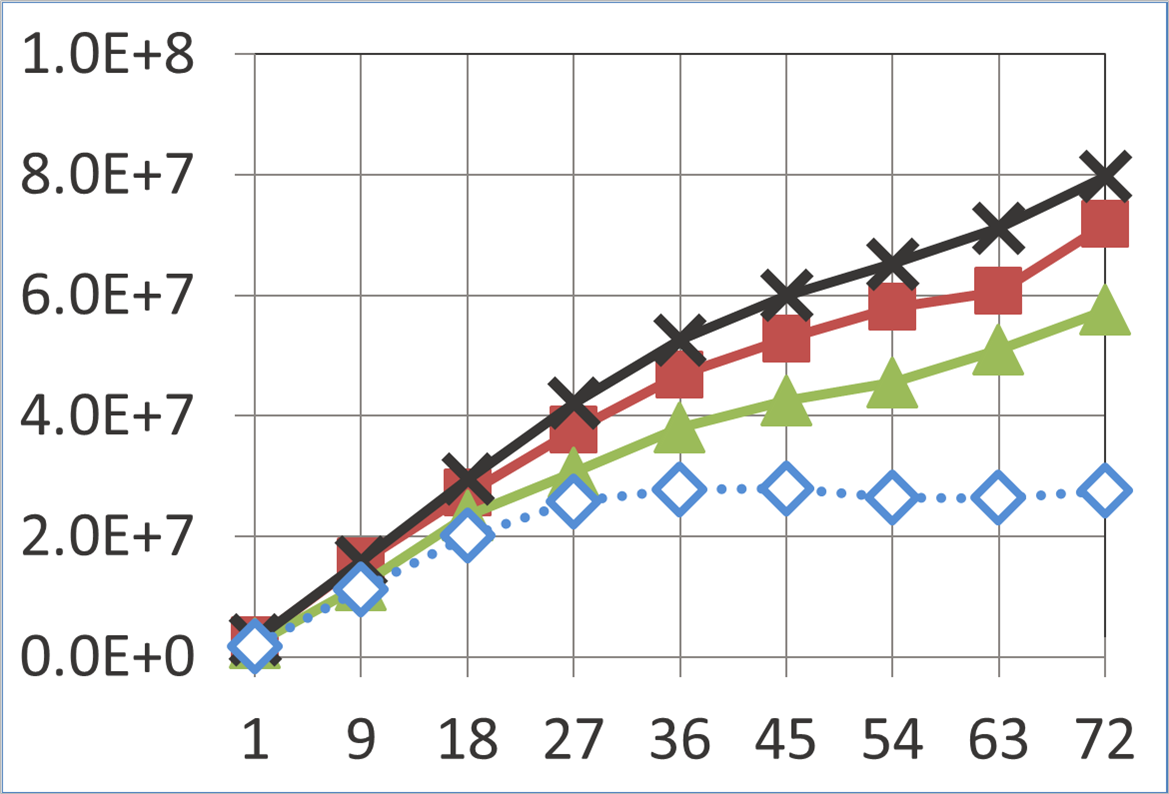} &
        \includegraphics[width=\linewidth]{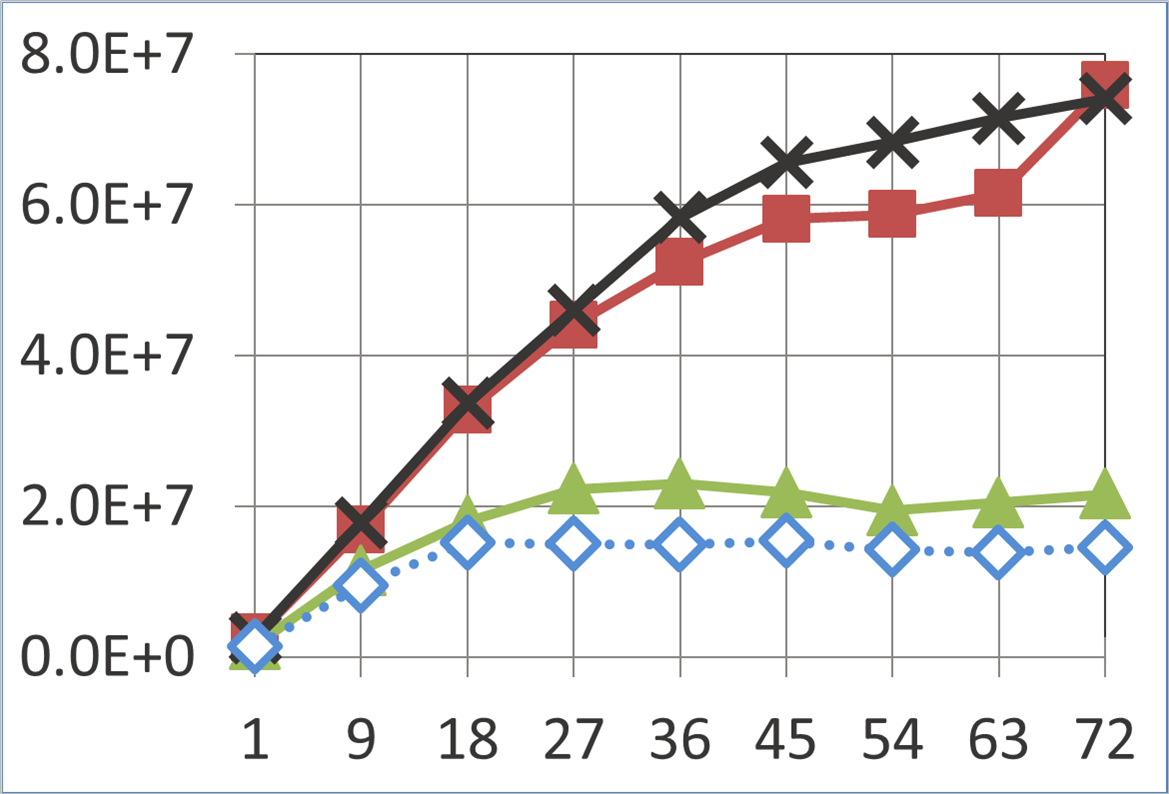}
        \\
        \vspace{-4.5mm}\rotatebox{90}{{\small\textbf{Heavy} workload}} &
        \vspace{-4.5mm}\includegraphics[width=\linewidth]{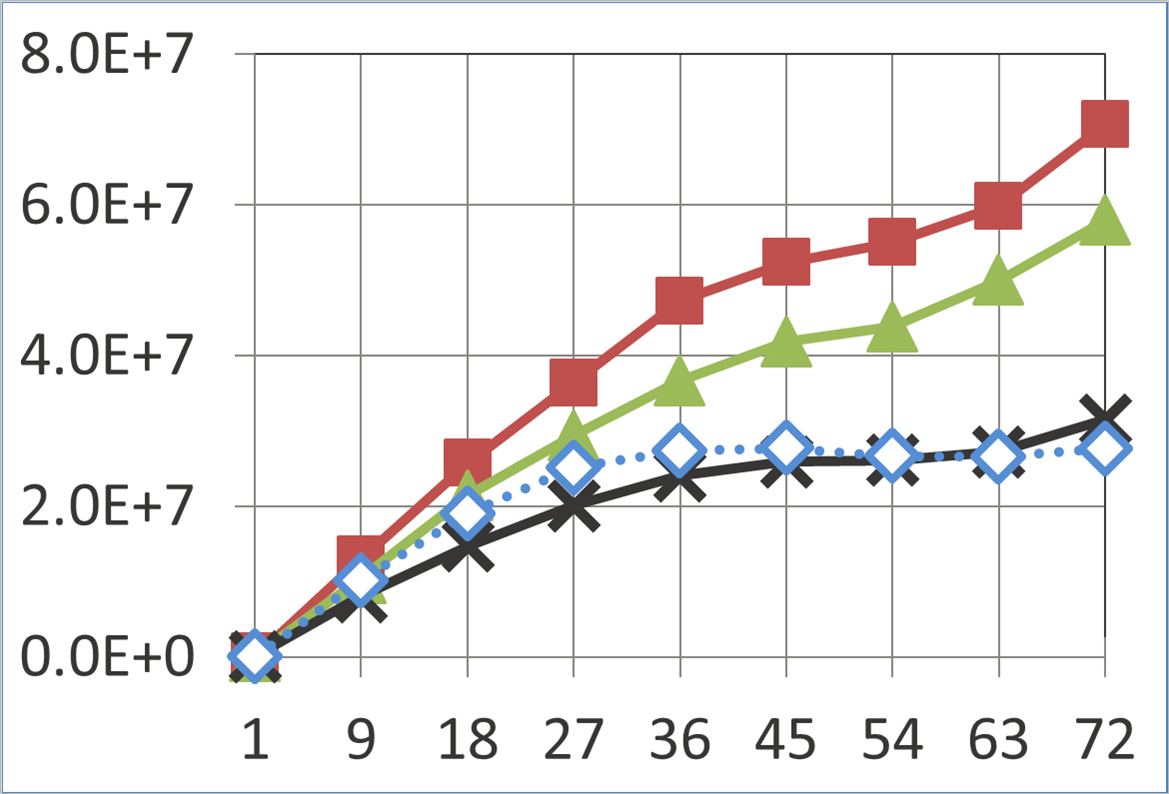} &
        \vspace{-4.5mm}\includegraphics[width=\linewidth]{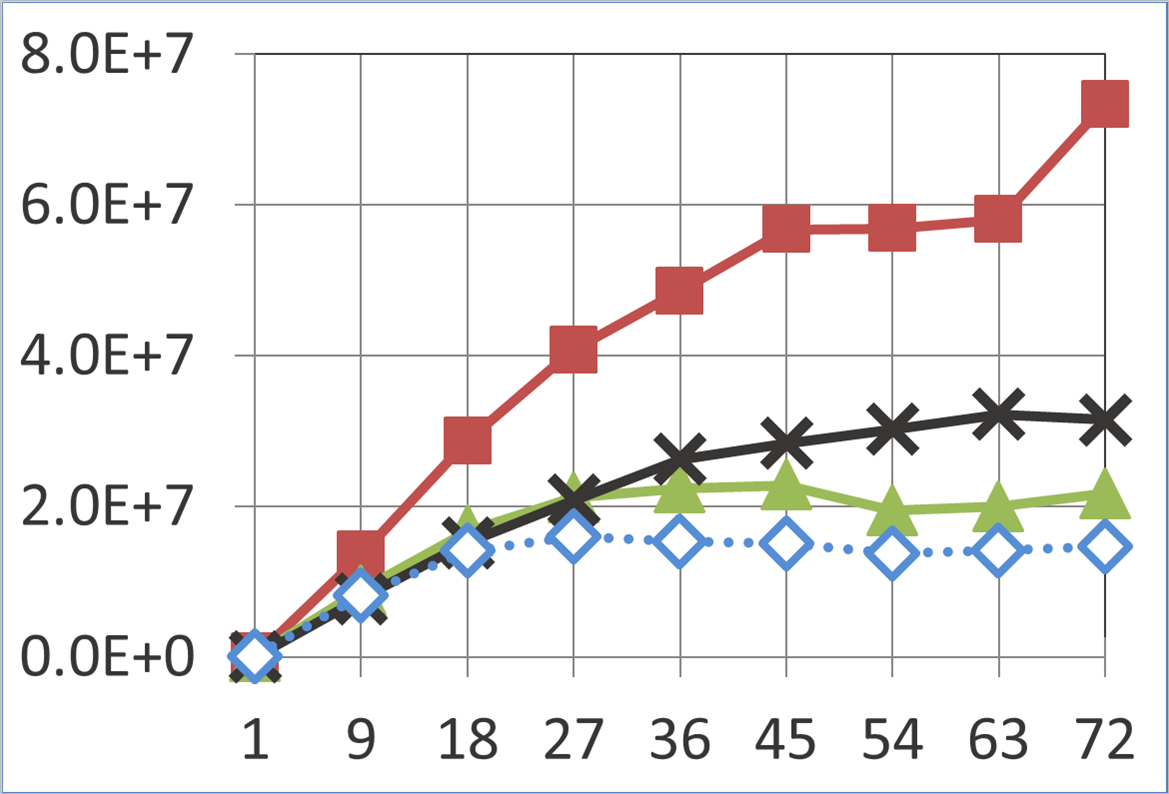}
        \\
    \end{tabular}
    
    \vspace{-1mm}
    \hspace{0.02\textwidth}\includegraphics[width=0.8\linewidth]{figures/perfgraphs/legendv4.png}
    \caption{Results (72 thread system) showing throughput (operations per second) vs. concurrent processes.}
    \label{fig-graphs2}
\end{figure}

Results from the 72-thread machine appear in Figure~\ref{fig-graphs2}. 
There, \textit{3-path} shows an even larger performance advantage over \textit{Non-HTM}.

\subsection{Code path usage and abort rates}
To gain further insight into the behaviour of our accelerated template implementations, we gathered some additional metrics about the experiments described above.
Here, we only describe results from the 48-thread Intel machine.
(Results from the 72-thread Intel machine were similar.)

\fakeparagraph{Operations completed on each path}
We started by measuring how often operations completed successfully on each execution path.
This revealed that operations almost always completed on the fast path.
Broadly, over all thread counts, the minimum number of operations completed on the fast path in any trial was 86\%, and the average over all trials was 97\%.

In each trial that we performed with 48 concurrent threads, at least 96\% of operations completed on the fast path, \textit{even in the workloads with RQs}.
Recall that RQs are the operations most likely to run on the fallback path, and they are only performed by a single thread, so they make up a relatively small fraction of the total operations performed in a trial.
In fact, our measurements showed that the number of operations which completed on the fallback path was never more than a fraction of one percent in our trials with 48 concurrent threads.

In light of this, it might be somewhat surprising that the performance of TLE was so much worse in heavy workloads than light ones.
However, the cost of serializing threads is high, and this cost is compounded by the fact that the operations which complete on the fallback path are often long-running.
Of course, in workloads where more operations run on the fallback path, the advantage of improving concurrency between paths would be even greater.


\begin{figure}[t]
    \centering
    \includegraphics[width=\linewidth]{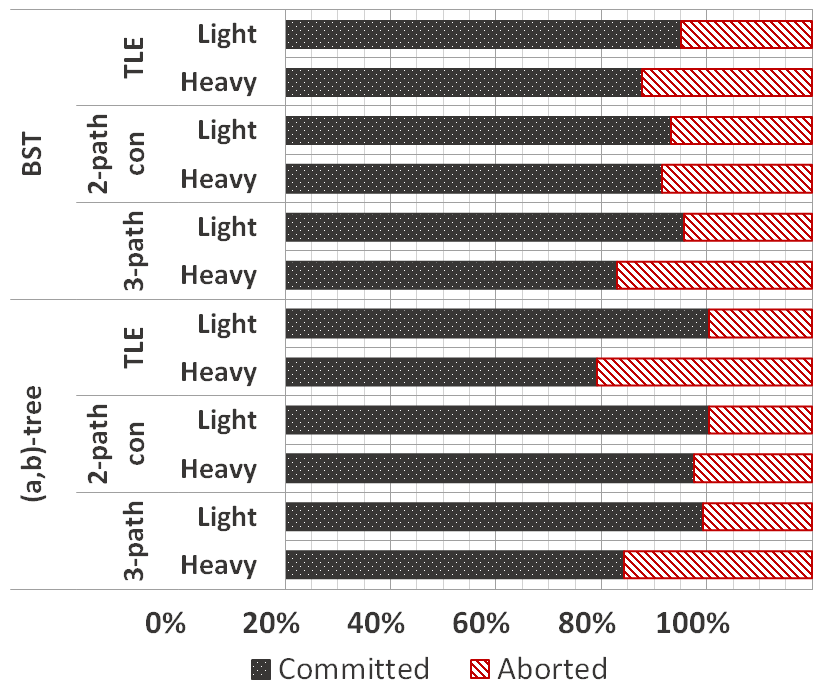} 
    \caption{How many transactions commit vs. how many abort in our experiments on the 48-thread machine.}
    \label{fig-abortrate}
\end{figure}

\fakeparagraph{Commit/abort rates}
We also measured how many transactions committed and how many aborted, on each execution path, in each of our trials.
Figure~\ref{fig-abortrate} summarizes the average commit/abort rates for each data structure, template implementation and workload.
Since nearly all operations completed on the fast path, we decided not to distinguish between the commit/abort rate on the fast path and the commit/abort rate on the middle path.

\subsection{Comparing with hybrid transactional memory}
\textit{Hybrid transactional memory} (hybrid TM) combines hardware and software transactions to hide the limitations of HTM and guarantee progress.
This offers an alternative way of using HTM to implement concurrent data structures.
Note, however, that state of the art hybrid TMs use locks.
So, \textbf{they cannot be used to implement lock-free data structures.}
Regardless, to get an idea of how such implementations would perform, relative to our accelerated template implementations, we implemented the unbalanced BST using Hybrid NOrec, which is arguably the fastest hybrid TM implementation with readily available code~\cite{hynorecriegel}.

If we were to use a precompiled library implementation of Hybrid NOrec, then the unbalanced BST algorithm would have to perform a library function call for \textit{each read and write to shared memory}, which would incur significant overhead.
So, we directly compiled the code for Hybrid NOrec into the code for the BST, allowing the compiler to inline the Hybrid NOrec functions for reading and writing from shared memory into our BST code, eliminating this overhead.
Of course, if one intended to use hybrid TM in practice (and not in a research prototype), one would use a precompiled library, with all of the requisite overhead.
Thus, the following results are quite charitable towards hybrid TMs.

\begin{figure}[t]
    \centering
    \begin{minipage}{\linewidth}
        \centering
        \setlength\tabcolsep{0pt}
        \begin{tabular}{m{0.03\linewidth}m{0.85\linewidth}}
            &
            \fcolorbox{black!80}{black!40}{\parbox{\dimexpr \linewidth-2\fboxsep-2\fboxrule}{\centering\textbf{2x 24-thread Intel E7-4830 v3}}}
            \\
            &
            \fcolorbox{black!50}{black!20}{\parbox{\dimexpr \linewidth-2\fboxsep-2\fboxrule}{\centering {Unbalanced BST (LLX/SCX)\\ Updates w/key range $[0,10^4$)}}} 
            \\
            \vspace{-3mm}\rotatebox{90}{{\textbf{Light} workload}} &
            \includegraphics[width=\linewidth]{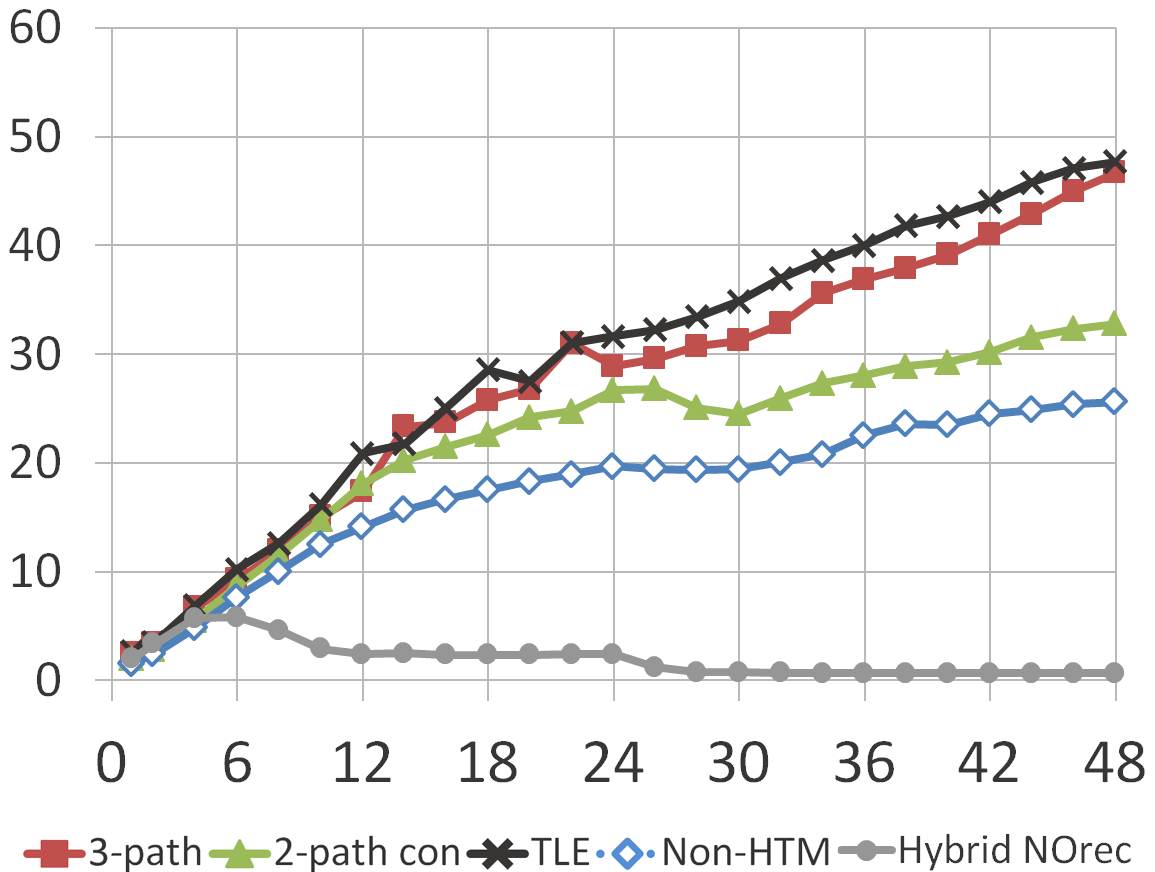} 
            \\
        \end{tabular}
        \caption{Results showing throughput (operations per second) versus number of processes for an unbalanced BST implemented with different tree update template algorithms, and with the hybrid TM algorithm \textit{Hybrid NOrec}.}
        \label{fig-graphs-hybridnorec}
    \end{minipage}
\end{figure}

We implemented the BST using Hybrid NOrec by wrapping sequential code for the BST operations in transactions, and manually replacing each read from (resp., write to) shared memory with a read (resp., write) operation provided by Hybrid NOrec.
Figure~\ref{fig-graphs-hybridnorec}
compares the performance of the resulting implementation to the other BST implementations discussed in Section~\ref{sec-3path-exp}.

The BST implemented with Hybrid NOrec performs relatively well with up to six processes.
However, beyond six processes, it experiences severe negative scaling.
The negative scaling occurs because Hybrid NOrec increments a global counter in each updating transaction (i.e., each transaction that performs at least one write).
This global contention hotspot in updating transactions causes many transactions to abort, simply because they contend on the global counter (and not because they conflict on any data in the tree).
However, even without this bottleneck, Hybrid NOrec would still perform poorly in heavy workloads, since it incurs very high instrumentation overhead for software transactions (which must acquire locks, perform repeated validation of read-sets, maintain numerous auxiliary data structures for read-sets and write-sets, and so on).
Note that this problem is not unique to Hybrid NOrec, as every hybrid TM must use a software TM as its fallback path in order to guarantee progress. 
In contrast, in our template implementations, the software-only fallback path is a fast lock-free algorithm.

\section{Modifications for performing searches outside of transactions} \label{sec-3path-search-outside}

In this section, we describe how the \textit{3-path} implementations of the unbalanced BST and relaxed ($a,b$)-tree can be modified so that each operation attempt on the fast path or middle path performs its search phase \textit{before} starting a transaction (and only performs its update phase in a transaction).
(The same technique also applies to the \textit{2-path con} implementations.)
First, note that the lock-free search procedure for each of these data structures is actually a standard, sequential search procedure.
Consequently, a simple sequential search procedure will return the correct result, regardless of whether it is performed inside a transaction.
(Generally, whenever we produce a \textit{3-path} implementation starting from a lock-free fallback path, we will have access to a correct non-transactional search procedure.)

The difficulty is that, when an operation starts a transaction and performs its update phase, it may be working on a part of the tree that was deleted by another operation.
One can imagine an operation $O_d$ that deletes an entire subtree, and an operation $O_i$ that inserts a node into that subtree.
If the search phase of $O_i$ is performed, then $O_d$ is performed, then the update phase of $O_i$ is performed, then $O_i$ may erroneously insert a node into the deleted subtree.

We fix this problem as follows.
Whenever an operation $O$ on the fast path or middle path removes a node from the tree, it sets a $marked$ bit in the node (just like operations on the fallback path do).
Whenever $O$ first accesses a node $u$ in its transaction, it checks whether $u$ has its $marked$ bit set, and, if so, aborts immediately.
This way, $O$'s transaction will commit only if every node that it accessed is in the tree.

We found that this modification yielded small performance improvements (on the order of 5-10\%) in our experiments.
The reason this improves performance is that fewer memory locations are tracked by the HTM system, which results in fewer capacity aborts.
We briefly discuss why the performance benefit is small in our experiments.
The relaxed ($a,b$)-tree has a very small height, because it is balanced, and its nodes contain many keys.
The BST also has a fairly small height (although it is considerably taller than the relaxed ($a,b$)-tree), because processes in our experiments perform insertions and deletions on uniformly random keys, which leads to trees of logarithmic height with high probability.
So, in each case, the sequence of nodes visited by searches is relatively small, and is fairly unlikely to cause capacity aborts.

The performance benefit associated with this modification will be greater for data structures, operations or workloads in which an operation's search phase will access a large number of nodes.
Additionally, IBM's HTM implementation in their POWER8 processors is far more prone to capacity aborts than Intel's implementation, since a transaction \textit{will} abort if it accesses more than 64 different cache lines~\cite{nguyen2015investigation}.
(In contrast, in Intel's implementation, a transaction can potentially commit after accessing tens of thousands of cache lines.)
Thus, this modification could lead to significantly better performance on POWER8 processors.

\section{More efficient memory reclamation on the fast path} \label{sec-3path-memrecl}

For the data structures presented in the paper, we implemented memory reclamation using an epoch based reclamation scheme called DEBRA~\cite{Brown:2015}.
This reclamation scheme is designed to reclaim memory for lock-free data structures, which are notoriously difficult to reclaim memory for. 
Since processes do not lock nodes before accessing them, one cannot simply invoke \texttt{free()} to release a node's memory back to the operating system as soon as the node is removed from the data structure.
This is because a process can always be poised to access the node just after it is freed.
The penalty for accessing a freed node is a program crash (due to a segmentation fault).
Thus, reclamation schemes like DEBRA implement special mechanisms to determine when it is safe to free a node that has been removed from the data structure.

However, advanced memory reclamation schemes become unnecessary if \textit{all} accesses to nodes are performed inside transactions.
With Intel's HTM, accessing freed memory inside a transaction cannot cause a segmentation fault and crash the program.
Instead, the transaction simply aborts. 
(Note, however, that this is not true for IBM's transactional memory implementation in their POWER8 processors.)
Consider a graph-based data structure whose operations are performed entirely in transactions.
In such a data structure, deleting and immediately freeing a node will simply cause any concurrent transaction that accesses the node (after it is freed) to abort.
This is because removing the node will change a pointer that was traversed during any concurrent search that reached the node.
Consequently, in such a data structure, reclaiming memory is as easy as invoking \texttt{free()} immediately after a node is removed.

In our three path algorithms, the fast path can only run concurrently with the middle path (but not the fallback path).
Thus, if every operation on the fast path or middle path runs entirely inside a transaction, then memory can be reclaimed on the fast path simply by using \texttt{free()} immediately after removing a node inside a transaction. 
Our performance experiments did \textit{not} implement this optimization, but doing so would likely further improve the performance of the three path algorithms. 

%
%
%

\section{Other uses for the \textit{3-path} approach} \label{sec-3path-other-uses}

\subsection{Accelerating data structures that use read-copy-update (RCU)}

In this section, we sketch a \textit{3-path} algorithm for an ordered dictionary implemented with a node-oriented unbalanced BST that uses the RCU synchronization primitives.
The intention is for this to serve as an example of how one might use the \textit{3-path} approach to accelerate a data structure that uses RCU.

RCU is both a programming paradigm and a set of synchronization primitives.
The paradigm organizes operations into a search/reader phase and an (optional) update phase.
In the update phase, all modifications are made on a \textit{new copy} of the data, and the old data is atomically replaced with the new copy.
In this work, we are interested in the \textit{RCU primitives} (rather than the paradigm).

\textbf{Semantics of RCU primitives and their uses.}
The basic RCU synchronization primitives are \textit{rcu\_begin}, \textit{rcu\_end} and \textit{rcu\_wait}~\cite{Desnoyers:2012}.
Operations invoke \textit{rcu\_begin} and \textit{rcu\_end} at the beginning and end of the search phase, respectively.
The interval between an invocation of \textit{rcu\_begin} and the next invocation of \textit{rcu\_end} by the same operation is called a \textit{read-side critical section}.
An invocation of \textit{rcu\_wait} blocks until all read-side critical sections that started before the invocation of \textit{rcu\_wait} have ended.
%
One common use of \textit{rcu\_wait} is to wait, after a node has been deleted, until no readers can have a pointer to it, so that it can safely be freed.
It is possible to use RCU as the sole synchronization mechanism for an algorithm if one is satisfied with allowing many concurrent readers, but only a single updater at a time.
If multiple concurrent updaters are required, then another synchronization mechanism, such as fine-grained locks, must also be used.
However, one must be careful when using locks with RCU, since locks cannot be acquired inside a read-side critical section without risking deadlock.

\textbf{The CITRUS data structure.}
We consider how one might accelerate a node-oriented BST called CITRUS~\cite{arbel2014concurrent}, which uses the RCU primitives, and fine-grained locking, to synchronize between threads.
First, we briefly describe the implementation of CITRUS.
At a high level, RCU is used to allow operations to search without locking, and fine-grained locking is used to allow multiple updaters to proceed concurrently.

The main challenge in the implementation of CITRUS is to prevent race conditions between searches (which do not acquire locks) and deletions.
When an internal node $u$ with two children is deleted in an internal BST, its key is replaced by its successor's key, and the successor (which is a leaf) is then deleted. 
This case must be handled carefully, or else the following can happen.
Consider concurrent invocations $D$ of \textit{Delete}($key$) and $S$ of \textit{Search}($key'$), where $key'$ is the successor of $key$.
Suppose $S$ traverses past the node $u$ containing $key$, and then $D$ replaces $u$'s key by $key'$, and deletes the node containing $key'$.
The search will then be unable to find $key'$, even though it has been in the tree throughout the entire search.
To avoid this problem in CITRUS, rather than changing the key of $u$ directly, $D$ replaces $u$ with a new copy that contains $key'$.
After replacing $u$, $D$ invokes \textit{rcu\_wait} to wait for any ongoing searches to finish, before finally deleting the leaf containing $key'$.
The primary sources of overhead in this algorithm are invocations of \textit{rcu\_wait}, and lock acquisition costs.

\textbf{Fallback path.}
The fallback path uses the implementation of CITRUS in~\cite{arbel2014concurrent} (additionally incrementing and decrementing the gobal fetch-and-add object $F$, as described in Section~\ref{sec-3path-algs}).

\textbf{Middle path.}
The middle path is obtained from the fallback path by wrapping each fallback path operation in a transaction and optimizing the resulting code.
The most significant optimization comes from an observation that the invocation of \textit{rcu\_wait} in \textit{Delete} is unnecessary since transactions make the operation atomic.
Invocations of \textit{rcu\_wait} are the dominating performance bottleneck in CITRUS, so this optimization greatly improves performance.
A smaller improvement comes from the fact that transactions can avoid acquiring locks.
Transactions on the middle path must ensure that all objects they access are not locked by other operations (on the fallback path), or else they might modify objects locked by operations on the fallback path. 
However, it is not necessary for transaction to actually \textit{acquire} locks.
Instead, it suffices for a transaction to simply \textit{read} the lock state for all objects it accesses (before accessing them) and ensure that they are not held by another process.
This is because transactions \textit{subscribe} to each memory location they access, and, if the value of the location (in this case, the lock state) changes, then the transaction will abort.

\textbf{Fast path.}
The fast path is a sequential implementation of a node-oriented BST whose operations are executed in transactions.
As in the other 3-path algorithms, each transaction starts by reading $F$, and aborts if it is nonzero.
This prevents operations on the fast path and fallback path from running concurrently.
There are two main differences between fast path and the middle path.
First, the fast path does not invoke \textit{rcu\_begin} and \textit{rcu\_end}.
These invocations are unnecessary, because operations on the fast path can run concurrently \textit{only} with other operations on the fast path or middle path, and \textit{neither} path depends on RCU for its correctness.
(However, the middle path \textit{must} invoke these operations, because it runs concurrently with the fallback path, which relies on RCU.)
The second difference is that the fast path does not need to read the lock state for any objects.
Any conflicts between operations on the fast and middle path are resolved directly by the HTM system. 

\subsection{Accelerating data structures that use k-CAS}

In this section, we sketch a \textit{3-path} algorithm for an ordered dictionary implemented with a singly-linked list that uses the $k$-CAS synchronization primitive.
The intention is for this to serve as an example of how one might use the \textit{3-path} approach to accelerate a data structure that uses $k$-CAS.

A $k$-CAS operation takes, as its arguments, $k$ memory locations, expected values and new values, and atomically: reads the memory locations and, if they contain their expected values, writes new values into each of them.
$k$-CAS has been implemented from single-word CAS~\cite{Harris:2002}.
We briefly describe this $k$-CAS implementation.
At a high level, a $k$-CAS creates a descriptor object that describes the $k$-CAS operation, then uses CAS to store a pointer to this descriptor in each memory location that it operates on.
Then, it uses CAS to change each memory location to contain its new value.
While a $k$-CAS is in progress, some fields may contain pointers to descriptor objects, instead of their regular values.
Consequently, reading a memory location becomes more complicated: it requires reading the location, then testing whether it contains a pointer to a descriptor object, and, if so, helping the $k$-CAS operation that it represents, before finally returning a value.

\textbf{Fallback path.}
The fallback path consists of the lock-free singly-linked list in~\cite{timnat2015practical}.
At a high level, each operation on the fallback path consists of a search phase, optionally followed by an update phase, which is performed using $k$-CAS.

\textbf{Middle path.}
Since the search phase in a linked list can be extremely long, and is likely to cause a transaction to abort (due to capacity limitations), the middle path was obtained by wrapping \textit{only the update phase} of each fallback path operation in a transaction, and optimizing the resulting code.
The main optimization on the middle path comes from replacing the software implementation of $k$-CAS with straightforward implementation from HTM (using the approach in~\cite{timnat2015practical}).
This HTM-based implementation performs the entire $k$-CAS atomically, so it does not need to create a descriptor, or store pointers to descriptors at nodes.

\textbf{Fast path.}
The fast path is a sequential implementation in which the \textit{update phase} of each operation is wrapped in a transaction.
The main optimization on the fast path comes from the fact that, since there are no concurrent operations on the fallback path, there are no $k$-CAS descriptors in shared memory.
Consequently, operations on the fast path do not need to check whether any values they read from shared memory are actually pointers to $k$-CAS descriptors, which can significantly reduce overhead.

\textbf{Preventing fast/fallback concurrency.}
Observe that our fast path optimization (to avoid checking whether any values that are read are actually pointers to $k$-CAS descriptors) is correct only if the search phase in the fast path does not run concurrently with the update phase of any operation on the fallback path.
For each of the other data structures we described, each operation runs entirely inside a single transaction.
Thus, for these data structures, it suffices to verify that the global fetch-and-add object $F$ is zero at the beginning of each transaction to guarantee that operations on the fast path do not run concurrently with operations on the fallback path.
However, this is not sufficient for the list, since only the update phase of each operation executes inside a transaction.
So, we need some extra mechanism to ensure that the fast path does not run concurrently with the fallback path.

If it is not important for the algorithm to be lock-free, then one can simply use a fast form of group mutual exclusion that allows many operations on the fast path, or many operations on the fallback path, but not both.
Otherwise, one can solve this problem by splitting the traversal into many small transactions, and verifying that $F$ is zero at the beginning of each.
If $F$ ever becomes non-zero, then some transaction will abort, and the enclosing operation will also abort.
%

\section{Related work} \label{sec-3path-related}

Hybrid TMs share some similarities to our work, since they all feature multiple execution paths.
The first hybrid TM algorithms allowed HTM and STM transactions to run concurrently~\cite{Kumar2006,Damron2006}.
Hybrid NOrec~\cite{Dalessandro2011} and Reduced hardware NOrec~\cite{Matveev2014} are hybrid TMs that both use global locks on the fallback path, eliminating any concurrency.
We discuss two additional hybrid TMs, Phased TM~\cite{Lev2007} (PhTM) and Invyswell~\cite{Calciu2014}, in more detail.

PhTM alternates between five \textit{phases}: HTM-only, STM-only, concurrent HTM/STM, 
and two global locking phases.
Roughly speaking, PhTM's HTM-only phase corresponds to our uninstrumented fast path, and its concurrent HTM/STM phase corresponds to our middle HTM and fallback paths.
However, their STM-only phase (which allows no concurrent hardware transactions) and global locking phases (which allow no concurrency) have no analogue in our approach.
In heavy workloads, 
PhTM must oscillate between its HTM-only and concurrent HTM/STM phases to maximize the performance benefit it gets from HTM.
When changing phases, PhTM typically waits until all in-progress transactions complete before allowing transactions to begin in the new mode.
Thus, after a phase change has begun, and before the next phase has begun, there is a window during which new transactions must wait (reducing performance).
One can also think of our three path approach as proceeding in two phases: one with concurrent fast/middle transactions and one with concurrent middle/fallback transactions. 
However, in our approach, ``phase changes'' do not introduce any waiting, and there is always concurrency between two execution paths.


%


Invyswell is closest to our three path approach.
At a high level, it features an HTM middle path and STM slow path that can run concurrently (sometimes), and an HTM fast path that can run concurrently with the middle path (sometimes) but not the slow path, and two global locking fallback paths (that prevent any concurrency).
Invyswell is more complicated than our approach, and has numerous restrictions on when transactions can run concurrently.
Our three path methodology does not have these restrictions.
The HTM fast path also uses an optimization called lazy subscription.
It has been shown that lazy subscription can cause opacity to be violated, which can lead to data corruption or program crashes~\cite{dice2014pitfalls}.

Hybrid TM is very general, and it pays for its generality with high overhead.
Consequently, data structure designers can extract far better performance for library code by using more specialized techniques.
Additionally, we stress that state of the art hybrid TMs use locks, so they cannot be used in lock-free data structures.

Different options for concurrency have recently begun to be explored in the context of TLE.
Refined TLE~\cite{dicerefined} and Amalgamated TLE~\cite{afek2015amalgamated} both improve the concurrency of TLE when a process is on the fallback path by allowing HTM transactions to run concurrently with a \textit{single process} on the fallback path.
Both of these approaches still serialize processes on the fallback path.
They also use locks, so they cannot be used to produce lock-free data structures.


Timnat, Herlihy and Petrank~\cite{timnat2015practical} proposed using a strong synchronization primitive called \textit{multiword compare-and-swap} ($k$-CAS) to obtain fast HTM algorithms.
They showed how to take an algorithm implemented using $k$-CAS and produce a two-path implementation that allows concurrency between the fast and fallback paths.
One of their approaches used a lock-free implementation of $k$-CAS on the fallback path, and an HTM-based implementation of $k$-CAS on the fast path.
They also experimented with two-path implementations that do not allow concurrency between paths, and found that allowing concurrency between the fast path and fallback path introduced significiteant overhead.
Makreshanski, Levandoski and Stutsman~\cite{makreshanski2015lock} also independently proposed using HTM-based $k$-CAS 
in the context of databases.

Liu, Zhou and Spear~\cite{Liu2015} proposed a methodology for accelerating concurrent data structures using HTM, and demonstrated it on several lock-free data structures.
Their methodology uses an HTM-based fast path and a non-transactional fallback path. 
The fast path implementation of an operation is obtained by encapsulating part (or all) of the operation in a transaction, and then applying sequential optimizations to the transactional code to improve performance.
Since the optimizations do not change the code's logic, the resulting fast path implements the same logic as the fallback path, so both paths can run concurrently.
Consequently, the fallback path imposes overhead on the fast path. 

Some of the optimizations presented in that paper are similar to some optimizations in our HTM-based implementation of \llt\ and \sct.
For instance, when they applied their methodology to the lock-free unbalanced BST of Ellen et~al.~\cite{Ellen:2010}, they observed that helping can be avoided on the fast path, and that the descriptors which are normally created to facilitate helping can be replaced by a small number of statically allocated descriptors.
However, they did not give details on exactly how these optimizations work, and did not give correctness arguments for them.
In contrast, our optimizations are applied to a more complex algorithm, and are proved correct.


Multiversion concurrency control (MVCC) is another way to implement range queries efficiently \cite{bernstein1983multiversion, attiya2012single}.
At a high level, it involves maintaining multiple copies of data to allow read-only transactions to see a consistent view of memory and serialize even in the presence of concurrent modifications. 
However, our approach could also be applied to operations that \textit{modify} a range of keys, so it is more general than MVCC. 


\section{Concluding remarks} \label{sec-3path-conclusion}


In this work, we explored the design space for HTM-based implementations of the tree update template of Brown et~al. and presented four accelerated implementations.
We discussed performance issues affecting HTM-based algorithms with two execution paths, and developed an approach that avoids them by using three paths.
We used our template implementations to accelerate two different lock-free data structures, and performed experiments that showed significant performance improvements over several different workloads.
This makes our implementations an attractive option for producing fast concurrent data structures for inclusion in libraries, where performance is critical.

Our accelerated data structures each perform an entire operation inside a single transaction (except on the fallback code path, where no transactions are used).
We discussed how one can improve efficiency by performing the read-only \textit{searching} part of an operation non-transactionally, and simply using a transaction to perform any modifications to the data structure.
Our \textit{3-path} approach may also have other uses.
As an example, we sketched an accelerated \textit{3-path} implementation of a node-oriented BST that uses the read-copy-update (RCU) synchronization primitives.
We suspect that a similar approach could be used to accelerate other data structures that use RCU.
Additionally, we described how one might produce a \textit{3-path} implementation of a lock-free algorithm that uses the $k$-CAS synchronization primitive.

\bibliographystyle{abbrv}
\bibliography{bibliography}

\end{document}

%% file: chap-template/tree-fig-1.pdf_t
\begin{picture}(0,0)%
\includegraphics{chap-template/tree-fig-1.pdf}%
\end{picture}%
\setlength{\unitlength}{2565sp}%
\begingroup\makeatletter\ifx\SetFigFont\undefined%
\gdef\SetFigFont#1#2#3#4#5{%
  \reset@font\fontsize{#1}{#2pt}%
  \fontfamily{#3}\fontseries{#4}\fontshape{#5}%
  \selectfont}%
\fi\endgroup%
\begin{picture}(5925,3003)(4441,-5902)
\put(10351,-4411){\makebox(0,0)[lb]{\smash{{\SetFigFont{7}{8.4}{\rmdefault}{\mddefault}{\updefault}{\color[rgb]{0,0,0}$N$}%
}}}}
\put(4501,-4486){\makebox(0,0)[lb]{\smash{{\SetFigFont{7}{8.4}{\rmdefault}{\mddefault}{\updefault}{\color[rgb]{0,0,0}$R$}%
}}}}
\put(5326,-3061){\makebox(0,0)[lb]{\smash{{\SetFigFont{7}{8.4}{\rmdefault}{\mddefault}{\updefault}{\color[rgb]{0,0,0}$parent$}%
}}}}
\put(8326,-3061){\makebox(0,0)[lb]{\smash{{\SetFigFont{7}{8.4}{\rmdefault}{\mddefault}{\updefault}{\color[rgb]{0,0,0}$parent$}%
}}}}
\put(7501,-4486){\makebox(0,0)[lb]{\smash{{\SetFigFont{7}{8.4}{\rmdefault}{\mddefault}{\updefault}{\color[rgb]{0,0,0}$R$}%
}}}}
\end{picture}%